\documentclass[%
    reprint,
    superscriptaddress,
    amsmath,
    amssymb,
    aps,
]{revtex4-2}

\usepackage{graphicx}

\usepackage{dcolumn}
\usepackage{bm}
\usepackage{subfiles}
\usepackage[
    acronym, toc
]{glossaries}
\usepackage[ampersand]{easylist}
\usepackage{float}
\usepackage{xcolor}
\usepackage{amsfonts}
\usepackage{braket}
\usepackage{enumitem}
\usepackage{lineno}
\usepackage{nicefrac}
\usepackage[toc,page,titletoc]{appendix}
\usepackage{multirow}
\usepackage{natbib}

\usepackage[ruled]{algorithm2e} 
\usepackage{newfloat,algcompatible} 
\usepackage[pdftex,colorlinks=true,bookmarks=false,citecolor=blue,urlcolor=blue]{hyperref}

\hypersetup{
    colorlinks=true,
    linkcolor=blue,
    filecolor=magenta,      
    urlcolor=blue,
}
\usepackage{tikz}
\usetikzlibrary{shapes.geometric, arrows, positioning, fit,calc}

\usepackage[capitalise, nameinlink]{cleveref}
\makeglossaries

\DeclareMathOperator*{\argmax}{arg\,max}

\newcommand\ho{\hat{H}_0}
\newcommand\hi{\hat{H}_i}
\newcommand\hj{\hat{H}_j}
\newcommand\hp{\hat{H}^{\prime}}
\newcommand\ha{\hat{H}_a}
\newcommand\hb{\hat{H}_b}

\newcommand\huc{\hat{H}_{C}^{\mu}}

\newcommand\s{\hat{\sigma}}
\newcommand\sx{\hat{\sigma}^{x}}
\newcommand\sy{\hat{\sigma}^{y}}
\newcommand\sz{\hat{\sigma}^{z}}
\newcommand\al{\vec{\alpha}}

\newcommand\lk{ l } 
\newcommand\tll{{\mathcal{L}}} 

\newcommand{\expset}{{ \mathcal{E} }}
\newcommand{\expdata}{{ \mathcal{D} }}
\newcommand{\population}{\mathcal{P}}
\newcommand{\termset}{{ \mathcal{T} }}
\newcommand{\termseto}{{ \termset_0 }}
\newcommand{\termseti}{{ \termset_i }} 
\newcommand{\terms}{{ \vec{T} }}
\newcommand\To{{ \termset_0 }}
\newcommand\Ti{{ \termset_i }}
\newcommand{\fs}{ \ensuremath{ F_{1}\text{-score} }}
\newcommand\bij{B_{ij}}
\newcommand{\icol}[1]{
  \left(\begin{matrix}#1\end{matrix}\right)%
}
\newcommand{\irow}[1]{
  \begin{matrix}(#1)\end{matrix}%
}
\newcommand{\ttt}[1]{\texttt{#1}}

\begin{document}

    \author{Brian Flynn}
\email{brian.flynn.bristol@gmail.com}
\thanks{Present address: Phasecraft, UK}
\affiliation{Quantum Engineering Technology Labs, University of Bristol, BS8 1FD, Bristol, UK}
\affiliation{Quantum Engineering Centre for Doctoral Training, University of Bristol, Bristol BS8 1FD, UK}

\author{Antonio A. Gentile}
\email{antonio.a.gentile@hotmail.it}
\affiliation{Quantum Engineering Technology Labs, University of Bristol, BS8 1FD, Bristol, UK}
\affiliation{Qu \& Co BV, 1070 AW, Amsterdam, the Netherlands.}

\author{Nathan Wiebe}
\email{nwiebe@cs.toronto.edu}
\affiliation{Department of Computer Science, University of Toronto, Toronto, Canada}
\affiliation{Pacific Northwest National Laboratory, Richland, United States}
\affiliation{University of Washington, Seattle, United States}

\author{Raffaele Santagati} 
\email{raffaele.santagati@boehringer-ingelheim.com}
\affiliation{Quantum Engineering Technology Labs, University of Bristol, BS8 1FD, Bristol, UK}
\affiliation{Boehringer-Ingelheim Quantum Lab, Doktor-Boehringer-Gasse 5-11, 1120 Wien, Austria.}

\author{Anthony Laing}
\email{anthony.laing@bristol.ac.uk}
\affiliation{Quantum Engineering Technology Labs, University of Bristol, BS8 1FD, Bristol, UK}

    \clearpage 
    \title{
        Quantum Model Learning Agent: characterisation of quantum systems through machine learning
    }
    \date{\today}
    \newacronym{qmla}{QMLA}{quantum model learning agent}
\newacronym{qhl}{QHL}{quantum Hamiltonian learning}
\newacronym{bf}{BF}{Bayes factor}
\newacronym{nisq}{NISQ}{noisy intermediate scale quantum}
\newacronym{ml}{ML}{machine learning}
\newacronym{qml}{QML}{quantum machine learning}
\newacronym{nv}{NV}{nitrogen-vacancy}
\newacronym{ga}{GA}{genetic algorithm}
\newacronym{gr}{GR}{Growth Rule}
\newacronym[plural=ESs, firstplural=exploration strategies (ESs)]{es}{ES}{exploration strategy}
\newacronym{et}{ET}{exploration tree}
\newacronym{fh}{FH}{Fermi-Hubbard}
\newacronym{fhm}{FHM}{Fermi-Hubbard model}
\newacronym{im}{IM}{Ising model}
\newacronym{hm}{HM}{Heisenberg model}
\newacronym{bfeer}{BFEER}{Bayes factor enhanced Elo ratings}
\newacronym{ges}{GES}{genetic exploration strategy}
\newacronym{smc}{SMC}{sequential Monte Carlo}
\newacronym{tll}{TLL}{total log-likelihood}
\newacronym{aic}{AIC}{Akaike information criterion}
\newacronym{aicc}{AICC}{Akaike information criterion corrected}
\newacronym{bic}{BIC}{Bayesian information criterion}
\newacronym{ll}{LL}{log-likelihood}
\newacronym{of}{OF}{objective function}
\newacronym{tp}{TP}{true positives}
\newacronym{tn}{TN}{true negatives}
\newacronym{fp}{FP}{false positives}
\newacronym{fn}{FN}{false negatives}
\newacronym{iqle}{IQLE}{interactive quantum likelihood estimation}

    \begin{abstract}
Accurate models of real quantum systems are important for investigating their behaviour, yet are difficult to distill empirically. 
Here, we report an algorithm -- the Quantum Model Learning Agent (QMLA) -- to reverse engineer Hamiltonian descriptions of a target system. 
We test the performance of QMLA on a number of simulated experiments,  demonstrating several mechanisms for the design of candidate Hamiltonian models
and simultaneously entertaining  numerous hypotheses 
about the nature of the physical interactions governing the system under study.
QMLA is shown to identify the true model in the majority of instances, when provided with limited a priori information, and control of the experimental setup. 
Our protocol can explore Ising, Heisenberg and Hubbard families of models in parallel, reliably identifying the family which best describes the system dynamics.
We demonstrate QMLA operating on large model spaces by incorporating a genetic algorithm to formulate new hypothetical models.
The selection of models whose features propagate to the next generation is based upon an objective function inspired by the Elo rating scheme, typically used to rate competitors in games such as chess and football.
In all instances, our protocol finds models 
that exhibit $\fs \geq 0.88$ when compared with the true model, 
and it precisely identifies the true model in 72\% of cases,
whilst exploring a space of over $250,000$ potential models. 
By testing which interactions actually occur in the target system, QMLA is a viable tool for both the exploration of fundamental physics and the characterisation and calibration of quantum devices.
    \end{abstract}
    \maketitle

\section{Introduction}\label{sec:intro}


%

Efforts in the domain of quantum system characterisation and validation have encompassed \gls{ml} techniques~\cite{torlai2020machine}. 
\gls{ml} methodologies and statistical inference have found broad application in the wider development of quantum technologies, 
from error correction~\cite{chen2019machine, valenti2019hamiltonian} to nuclear magnetic resonance spectroscopy~\cite{obrien2021quantum},
and device calibration~\cite{Santagati2019Magnetic, Joas2021Online}.
Here we introduce an \gls{ml} algorithm which infers a model for quantum systems, 
allowing for automatic characterisation of such systems and devices. 

For a given black box quantum system, $Q$, its \emph{model} is the generator of its dynamics, e.g. its Hamiltonian $\ho$, consisting of a sum of independent terms, each of which correspond to a unique physical interaction contributing to $Q$'s dynamics. 
A growing set of quantum parameter estimation algorithms 
-- such as \gls{qhl}~\cite{Granade:2012kj, Ferrie:2012ip, wiebe2014qhlpra, bairey2019learning, evans2019scalable} 
among others~\cite{anshu2021sample, krastanov2019stochastic, wang2015hamiltonian, flurin2020using, niu2019learning, greplova2017quantum, lokhov2018optimal, acampora2019evolutionary, burgarth2017evolution, valenti2021scalable}
--
characterise quantum systems whose model is known in advance, by inferring an optimal parameterisation.

Leveraging parameter-learning as a subroutine, we introduce the \gls{qmla}, which aims to compose an approximate model $\hp$, by testing and comparing a series of candidate models against data drawn from $Q$.
\gls{qmla} differs from quantum parameter estimation algorithms  by removing the assumption of the precise form of $\ho$;
instead we use an optimisation routine to determine which terms ought to be included in $\hp$, 
thereby determining which interactions $Q$ is subject to.
\par 

\gls{qmla} was introduced in~\cite{gentile2020learning}, 
where an electron spin in a nitrogen-vacancy centre was characterised through its experimental measurements.
In this paper we generalise and extend \gls{qmla}, studying a wider class of theoretical systems to demonstrate the capability of the framework.
Firstly, we show that \gls{qmla} most often selects the best model, when presented with a pre-determined set of models, characterised by Ising, Heisenberg or Hubbard interactions, i.e. a number of distinct \emph{families}.
From a different perspective, we test \gls{qmla}'s ability to classify the family of model most appropriate to the target system. This illustrates a crucial application of \gls{qmla}, i.e. that it can be used to inspect the physical regime underlying a system of interest.
Finally we design a \gls{ga} within \gls{qmla} to explore a large \emph{model space}.
\par

The paper is organised as follows: in \cref{sec:qmla} we describe the \gls{qmla} protocol in detail; 
    in \cref{sec:case_studies} we describe several test cases for \gls{qmla} along with their results, 
    finishing with a discussion in \cref{sec:discussion}.

\section{Quantum Model Learning Agent}\label{sec:qmla}

\begin{figure*}[t]
    \centering
    \includegraphics{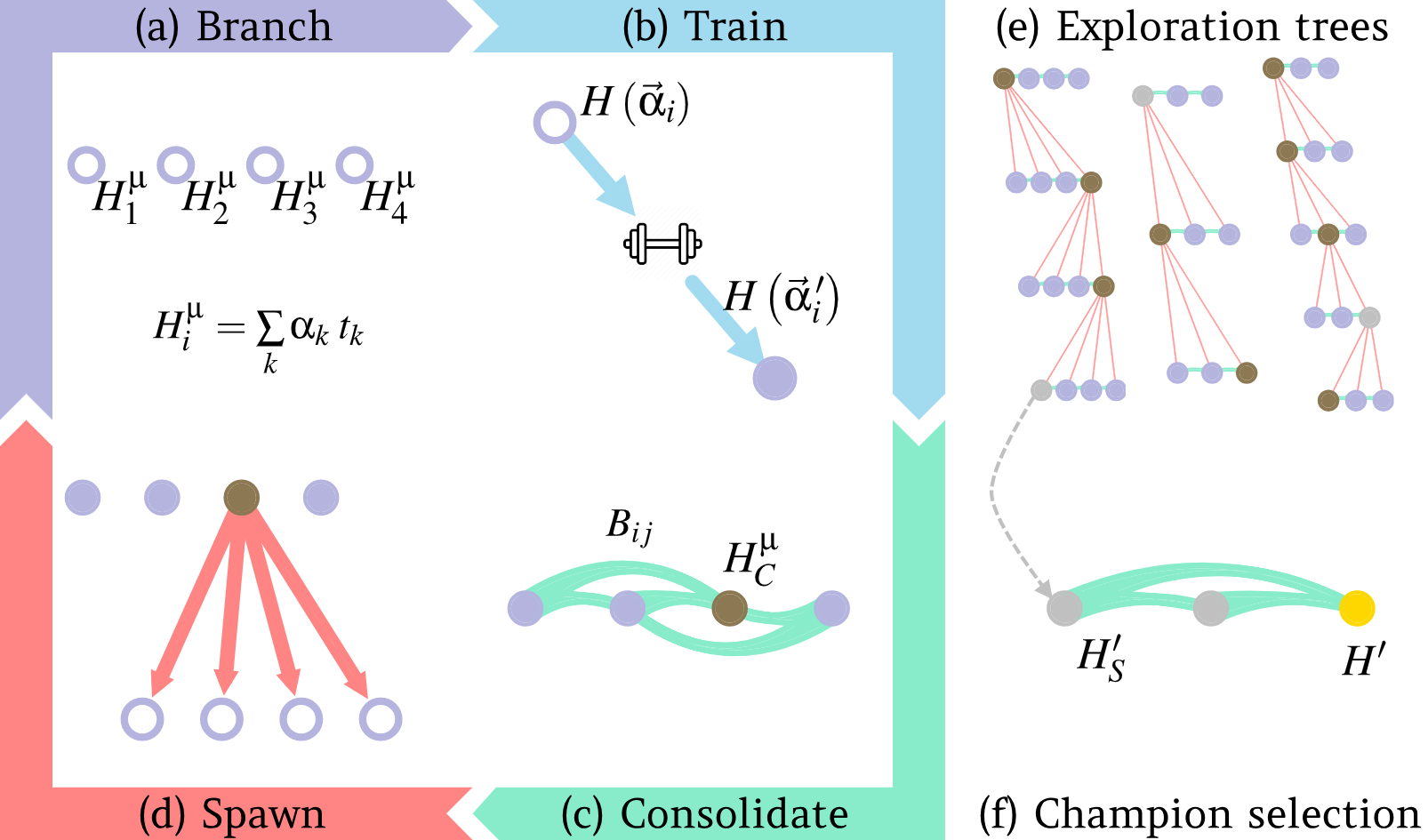}
    \caption{
        Schematic of \gls{qmla}. 
        \textbf{(a-d),} model search phase within an \glsentrylong{es}.
        \textbf{(a)}, Models are placed as (empty, purple) nodes on the \emph{active branch} $\mu$, 
            where each model is a sum of terms $\hat{t}_k$ multiplied by a scalar parameter $\alpha_k$. 
        \textbf{(b)}, Each active model is trained according to a subroutine, such as quantum Hamiltonian learning, to optimise $\al_i$, 
            resulting in the trained (filled, purple) node $\hat{H}(\al_i^{\prime})$. 
        \textbf{(c)}, $\mu$ is consolidated, i.e. models are evaluated relative to other
            models on $\mu$, according to the consolidation mechanism specificed by the \gls{es}.
            In this example, pairwise Bayes factors $B_{ij}$ between $\hi, \hj$ are computed, 
            resulting in the election of a single branch champion $\hat{H}_C^{\mu}$ (bronze). 
        \textbf{(d)}, A new set of models are \emph{spawned} according to the chosen
            \gls{es}'s model genertion strategy.
            In this example, models are spawned from a single parent. 
        \textbf{(e-f),} Higher level of entire QMLA procedure.
        \textbf{(e)}, The model search phase is presented on \emph{exploration trees}. 
            Multiple \gls{es} can operate in parallel, e.g. assuming different underlying physics.
            Each \gls{es} nominates a champion, $\hat{H}_{S}^{\prime}$ (silver), 
            after consolidating its branch champions (bronze). 
        \textbf{(f)}, $\hat{H}_{S}^{\prime}$ from each of the above exploration trees are gathered on a single branch, 
            which is consolidated to give the final champion model, $\hp$ (gold). 
    }
    \label{fig:qmla_schematic}
\end{figure*}

For a black-box quantum system, $Q$, whose Hamiltonian is given by $\ho$, 
\gls{qmla} distills a model $\hp \approx \ho$.
Models are characterised by their parameterisation and 
their constituent operators, or \emph{terms}. 
 For example, a model consisting of the sum of one-qubit Pauli terms, 
\begin{equation}
    \label{eqn:eg_model}
    \hat{H} 
        = \irow{\alpha_x & \alpha_y & \alpha_z} \cdot \icol{\sx \\ \sy \\ \sz} \\
        = \vec{\alpha} \cdot \terms,
\end{equation}
is characterised by its parameters $\vec{\alpha}$ and terms $\terms$.
\gls{qmla} aims primarily to find $\terms^{\prime}$, secondarily to find $\vec{\alpha}^{\prime}$, 
such that $\hp = \vec{\alpha}^{\prime} \cdot \terms^{\prime} \approx \ho$.
In doing so, $Q$ can be completely characterised.
\par

\gls{qmla} considers several \emph{branches} of candidate models, $\mu$: 
    we can envision individual models as leaves on a branch. 
Candidate models $\hi$ are \emph{trained} independently: 
    $\hi$ undergoes a parameter learning subroutine to optimise $\vec{\alpha}_i$, 
    under the assumption that $\hi=\ho$.  
In this work we use \gls{qhl} as the parameter learning subroutine~\cite{wiebe2014qhlpra}.  
While alternative parameter-learning methods can be used in principle, such as tomography~\cite{wang2015hamiltonian}, we focus on \gls{qhl} as it requires exponentially fewer samples than typical tomographic approaches.
Branches are \emph{consolidated}, meaning that models are evaluated relative to each other, 
    and ordered according to their strength. 
Models favoured by the consolidation are then used to \emph{spawn} a new set of models, 
    which are placed on the next branch;
    the average approximation of $\ho$ should thus improve with each new branch. 
\par

\emph{Exploration strategies} specify how \gls{qmla} can be used to target individual quantum systems, primarily by customising the consolidation and spawning mechanisms. Many possible heuristics can be proposed to explore the space; however, usually such methods choose the next branch of models is spawned by exploiting the knowledge gained during the previous branch.
For example, the single strongest model found in $\mu$ can be used as a basis for new models by adding a single further term from a prescribed set, i.e. \emph{greedy} spawning. 
The structure of \glspl{es} can be tuned to address systems' unique requirements;
    in later sections we outline the logic of \glspl{es} underlying the cases studied in this work. 
All aspects of \glspl{es} are explained in detail in \ref{sm:exploration_strategies}.
\par 

The QMLA procedure is depicted for an exemplary \gls{es} in \cref{fig:qmla_schematic}\textbf{a-d},
    centred on an iterative \emph{model search} phase as follows: 

\begin{easylist}[enumerate]
    & A set of models $\mathbb{H}^{\mu}$ are proposed (or spawned from a previous branch), and placed as leaves on branch $\mu$.
    & Train $\hat{H}_i \in \mathbb{H}^{\mu}$.
    && i.e. assuming $\ho = \hat{H}_i = \vec{\alpha}_i \cdot \terms_i$, optimise $\vec{\alpha}_i$.
    & Consolidate $\mu$
    && Evaluate and rank all $\hat{H}_i \in \mathbb{H}^{\mu}$.
    && Nominate the strongest model as the \emph{branch champion}, $\huc$.
    & $\mu \gets \mu + 1$
\end{easylist}

The model search phase is subject to termination criteria set by the \gls{es},
e.g. to terminate when a set number of branches is reached.
\gls{qmla} next consolidates the set of branch champions, $\{ \huc \}$, to declare the strongest model as the \emph{tree champion}, $\hat{H}_{S}^{\prime}$. 
Finally, \gls{qmla} can concurrently run multiple \gls{es}s, so the final step of \gls{qmla} is 
    to consolidate the set $\{ \hat{H}_{S}^{\prime} \}$, 
    in order to declare a \emph{global champion}, $\hp$. 
Each \gls{es} is assinged an independent \gls{et}, 
    consisting of as many branches as the \gls{es} requires. 
Together, then, we can think that the \gls{et}s constitute a forest, 
    such that \gls{qmla} can be described as a 
    \emph{forest search} for the single leaf (model) on any branch, 
    which best captures $Q$'s dynamics, 
    see \cref{fig:qmla_schematic}\textbf{(e,f)}.
\par

\section{Case Studies}\label{sec:case_studies}
\subsection{Model Selection}

\begin{figure*}[t]
    \centering
    \includegraphics[width=0.725\textwidth]{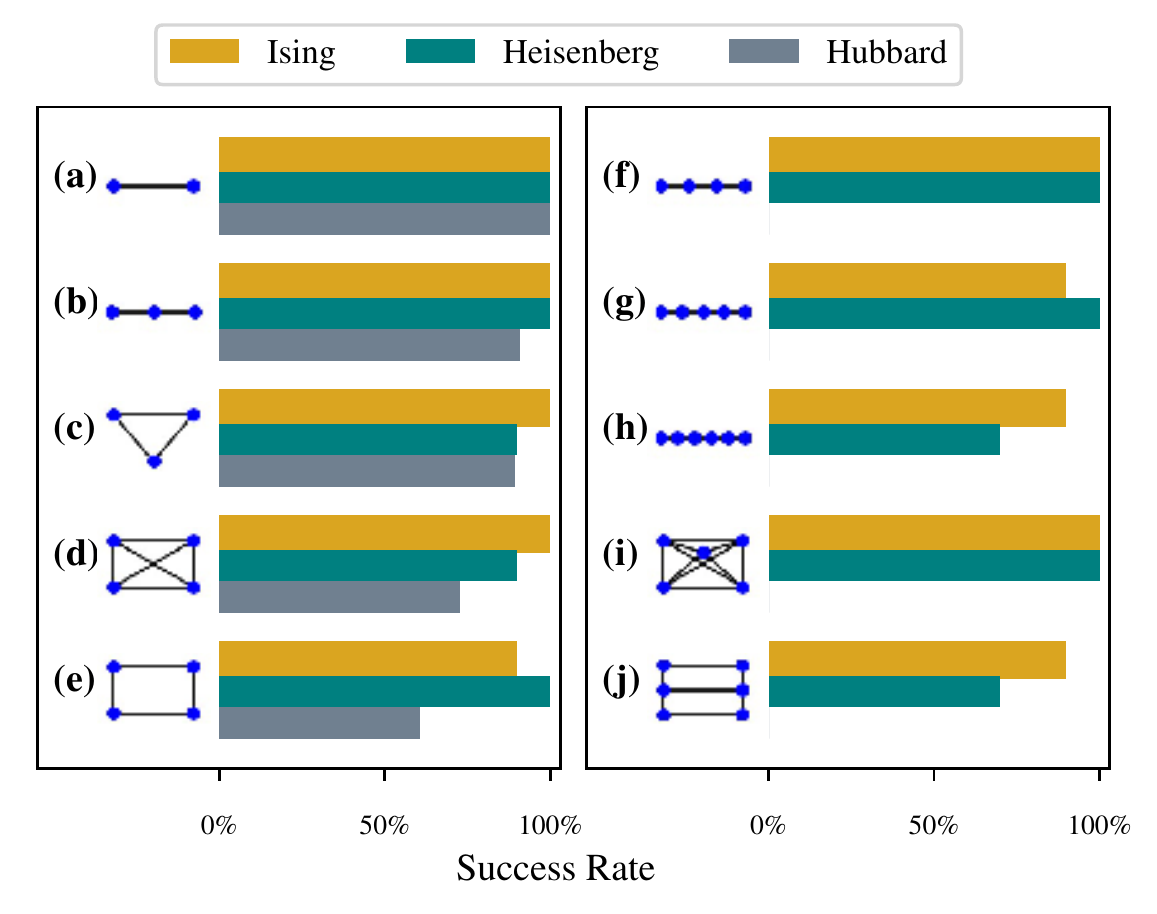}
    \caption{
        Rates of success for the quantum model learning agent (QMLA) in identifying the underlying lattice of the target model $\ho$.
        Each lattice (\textbf{a}-\textbf{e}) is set as the true model, $\ho$, for ten independent QMLA instances. 
        In each instance, QMLA considers the available lattices 
            (\textbf{a-j} for Ising and Heisenberg cases and \textbf{a-e} for the Hubbard case), 
            and selects a champion model, $\hp$, as the model most consistent with data generated by $\ho$. 
        The bars in each case show the percentage of instances for which QMLA identified $\hp=\ho$ when the corresponding lattice configuration gave $\ho$. 
    \label{fig:qmla_lattices}
    }
\end{figure*}

The simplest application of \gls{qmla} is to propose a fixed set of candidate models in advance:
    the \gls{et} is a single branch of prescribed models, with no further model spawning. 
The branch is consolidated as follows. 
Models are compared pairwise through \glspl{bf}, which provide statistical evidence of one model's superiority over the other~\cite{kass1995bayes}.
Each comparison gains a single point for the superior model; after all comparisons on the branch are complete, the model with most points is declared $\hp$. 
To demonstrate the reliability of \gls{qmla}, we begin with this case under three physical families, described by 
\textit{(i)} Ising; \textit{(ii)} Heisenberg and \textit{(iii)} Hubbard models.
\par

Candidate models differ in the number of lattice sites and the configuration of pairwise couplings, 
$\mathcal{C} = \{\langle k, l \rangle\}$.
\textit{(i)} is characterised by \crefrange{eqn:ising_params}{eqn:ising_terms}, 
likewise \textit{(ii)} is given by \crefrange{eqn:heisenberg_params}{eqn:heisenberg_terms},
and \textit{(iii)} by \crefrange{eqn:fh_params}{eqn:fh_terms}.

\begin{subequations}
    \begin{equation}
        \label{eqn:ising_params}
        \vec{\alpha}_I = \irow{ \alpha_{zz} & \alpha_{x} },
    \end{equation}
    \begin{equation}
        \label{eqn:ising_terms}
        \terms_I = \icol{ 
            \sum\limits_{\langle k, l \rangle \in \mathcal{C}} \s^{z}_{k} \s_{l}^{z} \\
            \sum\limits_{k=1}^{N} \s^{x}_{k}
        },
    \end{equation}

    \begin{equation}
        \label{eqn:heisenberg_params}
        \vec{\alpha}_H = \irow{ \alpha_{xx} & \alpha_{yy} & \alpha_{zz} },
    \end{equation}
    \begin{equation}
        \label{eqn:heisenberg_terms}
        \terms_{H} = \icol{ 
            \sum\limits_{\langle k, l \rangle \in \mathcal{C}} \s^{x}_{k} \s_{l}^{x} \\
            \sum\limits_{\langle k, l \rangle \in \mathcal{C}} \s^{y}_{k} \s_{l}^{y} \\
            \sum\limits_{\langle k, l \rangle \in \mathcal{C}} \s^{z}_{k} \s_{l}^{z}
        },
    \end{equation}

    \begin{equation}
        \label{eqn:fh_params}
        \vec{\alpha}_{FH} = \irow{ \alpha_{\uparrow} & \alpha_{\downarrow} & \alpha_{n} },
    \end{equation}
    \begin{equation}
        \label{eqn:fh_terms}
        \terms_{FH} = \icol{ 
            \sum\limits_{ \langle k, l \rangle \in \mathcal{C} }
                ( 
                    \hat{c}^{\dagger}_{l,\uparrow} \hat{c}_{k,\uparrow} + \hat{c}^{\dagger}_{k,\uparrow} \hat{c}_{l,\uparrow} 
                ) \\
            \sum\limits_{\langle k, l \rangle \in \mathcal{C}}
                ( 
                    \hat{c}^{\dagger}_{l,\downarrow} \hat{c}_{k,\downarrow} + \hat{c}^{\dagger}_{k,\downarrow} \hat{c}_{l,\downarrow} 
                ) \\
            \sum\limits_{k=1}^{N} \hat{n}_{k\uparrow} \hat{n}_{k\downarrow}
        },
    \end{equation}

\end{subequations}

where $\s_{k}^{p}\s_{l}^{p}$ is the coupling term between sites $k, l$ along axis $p$ (e.g. $\s_{k}^{x}\s_{l}^{x}$);
$N$ is the total number of sites in the system; 
$\hat{c}^{\dagger}_{l, s} \  (\hat{c}^{}_{l, s})$ 
is the fermionic creation (annihilation) operator for spin $s\in\{\uparrow, \downarrow\}$ on site $l$,
and $\hat{n}_{ks}$ is the on-site term for the number of spins of type  $s$ on site $k$. 
In order to simulate Hubbard Hamiltonians, we invoke a Jordan-Wigner transformation, resulting in a 2-qubit-per-site overhead, which renders simulations of this latter case more computationally demanding.
\par 

For each of \textit{(i)-(iii)}, we cycle through a set of lattice configurations $\mathbb{C} = \{\mathcal{C}_i\}$,
including 1D chains and 2D lattices with varying connectivity. 
$\| \mathbb{C} \| = 10$ for \textit{(i)} and \textit{(ii)}, 
and $\| \mathbb{C} \| = 5$ for \textit{(iii)} due to said simulation constraints.
The entertained lattices are shown in \cref{fig:qmla_lattices}, 
    along with the success rate with which \gls{qmla} finds precisely $\hp=\ho$ in instances where that lattice was set as $\ho$.
For every lattice type in each physical scenario, \gls{qmla} successfully identifies $\ho$ in $\geq60\%$ of instances, 
    proving the viability of \gls{bf} as a mechanism to distinguish models. 
\par

\begin{figure}[t!]
    \centering
    \includegraphics{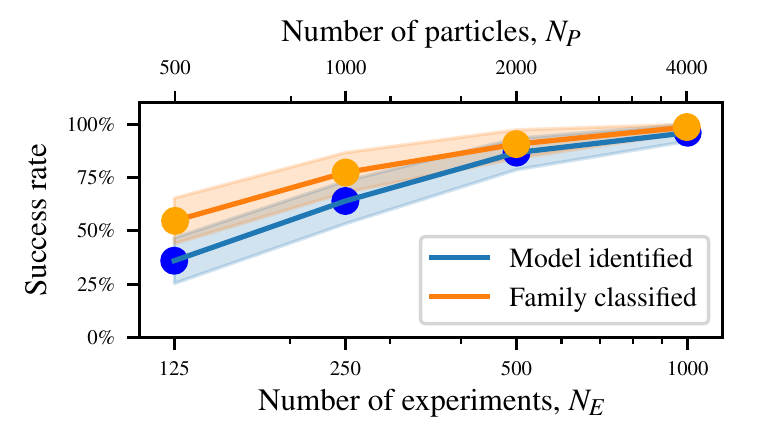}
    \caption{
    Rates of identification of the exact target model and its family with increasing training resources.
    The target model in each QMLA instance varies between Ising, Heisenberg and Hubbard families, set by lattices as in \cref{fig:qmla_lattices}.
    Each family is explored and nominates a $\hat{H}_{S}^{\prime}$; 
    $\hp$ is the champion of $\{ \hat{H}_{S}^{\prime}\}$, as in \cref{fig:qmla_schematic}.
    Each lattice per family is run as $\ho$ in five independent \gls{qmla} instances; 
    the success rate shows how often $\ho$ is identified (blue) and the family to which $\ho$ belongs is identified (orange) for increasing training resources $N_E, N_P$.
    }
    \label{fig:family_classiciation}
\end{figure}

Beyond identifying $\ho$ precisely, \gls{qmla} can distinguish which \emph{family} of model is most suitable for describing the target system, e.g. whether $Q$ is given by an Ising, Heisenberg or Hubbard model. 
We run an \gls{et} for the three named families in parallel; each nominates its tree champion $\hat{H}_{S}^{\prime}$, allowing \gls{qmla} to consolidate $\{\hat{H}_S^{\prime}\}$ to determine the champion model, $\hp$, and simultaneously the family to which $\hp$ belongs. 
Each model entertained by \gls{qmla} undergoes a training regime, the efficacy of which depends on the resources afforded, i.e. the number of experimental outcomes against which $\hi$ is trained, $N_E$, and the number of \emph{particles} used to approximate the parameters' probability distribution during Bayesian inference, $N_P$.
The concepts of experiments and particles are explained in \ref{sm:qhl}. 
The rate with which \gls{qmla} succeeds in identifying both $\ho$ and the family of $Q$ depends on these training resources, \cref{fig:family_classiciation}.

\subsection{Wider model search}
\renewcommand{\arraystretch}{2.0}
\setlength{\tabcolsep}{5pt}

\def\rowbox#1#2{%
    \smash{\color{#2}\fboxrule=1pt\relax\fboxsep=2pt\relax%
    \llap{\rlap{\fbox{\vphantom{0}\makebox[#1]{}}}~}}\ignorespaces
}

\definecolor{parentA}{HTML}{002bd8} 
\definecolor{parentB}{HTML}{316c00} 
\definecolor{mutation}{HTML}{961f29} 
\newcommand\parentcolourA{parentA}
\newcommand\parentcolourB{parentB}
\newcommand\mutationcolour{mutation}

\newcommand\longrowboxlenth{185pt}
\newcommand\shortrowboxlength{80pt}

\begin{table*}

    \begin{center}
    \begin{tabular}{ c  c | c  c  c  c  c  c } 
            \hline
        \multicolumn{2}{c|}{Model} & \multicolumn{6}{c}{Chromosome} \\
        & $\terms$ & $\hat{\sigma}_{(1,2)}^{x}$ & $\hat{\sigma}_{(1,2)}^{z}$ & $\hat{\sigma}_{(2,3)}^{y}$ 
            & $\hat{\sigma}_{(2,3)}^{x}$ & $\hat{\sigma}_{(2,3)}^{y}$ & $\hat{\sigma}_{(2,3)}^{x}$ 
            \\ 
        \hline
        \textcolor{\parentcolourA}{$\gamma_{p_1}$} & $\irow{ 
            \textcolor{\parentcolourA}{ \hat{\sigma}_{(1,2)}^{x}} 
            & \textcolor{\parentcolourA}{ \hat{\sigma}_{(1,2)}^{z}} 
            & \textcolor{\parentcolourA}{ \hat{\sigma}_{(2,3)}^{y}} 
        }$
        & \rowbox{\longrowboxlenth}{\parentcolourA} 1 & 0 & 1 & 0 & 1 & 0 \\
        \textcolor{\parentcolourB}{$\gamma_{p_2}$} & $\irow{ 
            \textcolor{\parentcolourB}{ \hat{\sigma}_{(1,2)}^{z}} 
            & \textcolor{\parentcolourB}{ \hat{\sigma}_{(2, 3)}^{y}} 
            & \textcolor{\parentcolourB}{ \hat{\sigma}_{(2,3)}^{z}} 
        }$
        & \rowbox{\longrowboxlenth}{\parentcolourB} 0 & 0 & 1 & 0 & 1 & 1 \\

        \hline
        $\gamma_{c_1}$ & $\irow{ 
            \textcolor{\parentcolourA}{ \hat{\sigma}_{(1,2)}^{x}} 
            & \textcolor{\parentcolourA}{ \hat{\sigma}_{(1,2)}^{z}} 
            & \textcolor{\parentcolourB}{ \hat{\sigma}_{(2, 3)}^{y}} 
            & \textcolor{\parentcolourB}{ \hat{\sigma}_{(2,3)}^{z}} 
        }$
        & \rowbox{\shortrowboxlength}{\parentcolourA} 1 & 0 & 1 & \rowbox{\shortrowboxlength}{\parentcolourB} 0 & 1 & 1 \\
        $\gamma_{c_2}$ & $\irow{ 
            \textcolor{\parentcolourA}{ \hat{\sigma}_{(1,2)}^{z}} 
            & \textcolor{\parentcolourB}{ \hat{\sigma}_{(2, 3)}^{y}} 
        }$
        & \rowbox{\shortrowboxlength}{\parentcolourB} 0 & 0 & 1 & \rowbox{\shortrowboxlength}{\parentcolourA} 0 & 1 & 0\\

        \hline

        $\gamma_{c_2}^{\prime}$ & $\irow{ 
            \textcolor{\parentcolourA}{ \hat{\sigma}_{(1,2)}^{z}} 
            & \textcolor{\mutationcolour}{ \hat{\sigma}_{(2, 3)}^{x}} 
            & \textcolor{\parentcolourB}{ \hat{\sigma}_{(2, 3)}^{y}} 
        }$
        & 0 & 0 & 1 & \rowbox{10pt}{\mutationcolour} 1 & 1 & 0\\
        \hline 
    \end{tabular}

    \caption[Mapping between \gls{qmla}'s models and chromosomes used by a genetic algorithm.]{
        Mapping between \gls{qmla}'s models and chromosomes used by a genetic algorithm. 
        Example shown for a three-qubit system with six possible terms, $\s_{i,j}^{w} = \s_i^w \s_j^w$. 
        Model terms are mapped to binary genes: 
            if the gene registers $0$ then the corresponding term is not present in the model, 
            and if it registers $1$ the term is included. 
        The top two chromosomes are \emph{parents}, $\gamma_{p_1}=101010$ (blue) and $\gamma_{p_2}=001011$ (green):
            they are mixed to spawn new models. 
        We use a one--point cross over about the midpoint:
            the first half of $\gamma_{p_1}$ is mixed with the second half of $\gamma_{p_2}$ 
            to produce two new children chromosomes, $\gamma_{c_1}, \gamma_{c_2}$. 
        Mutation occurs probabilistically: each gene has a 25$\%$ chance of being mutated, e.g. a single gene (red) flipping from $0 \rightarrow 1$ to mutate $\gamma_{c_2}$ to $\gamma_{c_2}^{\prime}$.
        The next generation of the genetic algorithm will then include $\gamma_{c_1}, \gamma_{c_2}^{\prime}$ (assuming $\gamma_{c_1}$ does not mutate). 
        To generate $N_m$ models for each generation, $N_m/2$ parent couples are sampled from the previous generation and crossed over. 
    }
    \label{table:chromosome_example}
    \end{center}
\end{table*}

\begin{figure*}[t!]
    \includegraphics[width=0.99\textwidth]{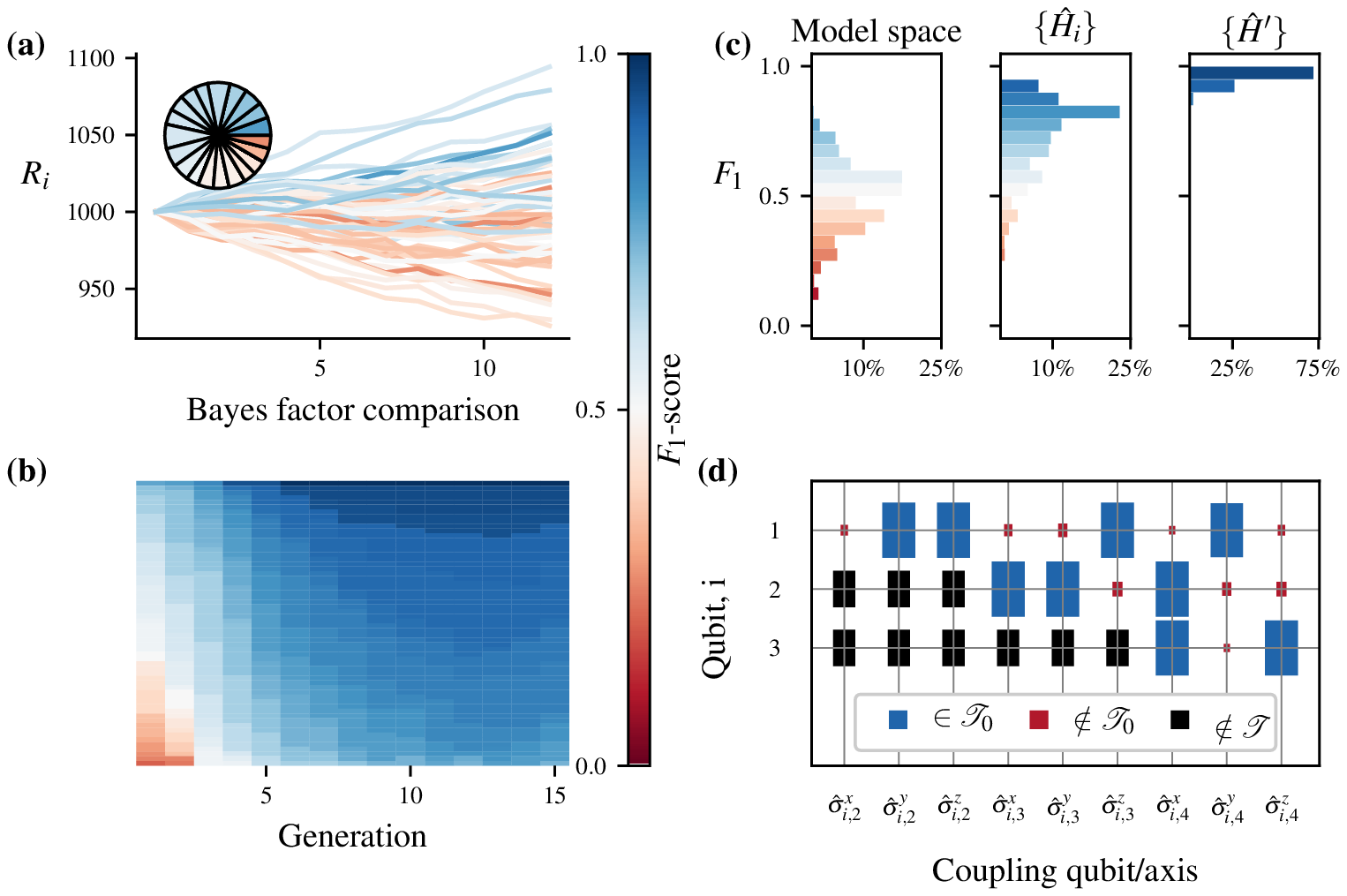}
    \caption{
        \label{fig:gen_alg_analysis}
        Genetic algorithm exploration strategy within QMLA. 
        \textbf{(a-b), Single instance of QMLA.} 
            The genetic algorithm runs for $N_g=15$ generations, where each generation tests $N_m=60$ models.
            \textbf{(a)}, Ratings of all models in a single \glsentrylong{ga} generation.
            Each line represents a unique model and is coloured by the $\fs$ of that model. 
            \textbf{Inset}, the selection probabilities resulting from the final ratings of this generation, 
                i.e. the models' chances of being chosen as a parent to a new model. 
            Only a fraction of models are assigned selection probability, while the remaining poorer-performing 
            models are truncated.     
            \textbf{(b)},  Gene pool progression for $N_m=60, N_g=15$. 
            Each tile at each generation represents a model by its $\fs$. 
        \textbf{(c-d), Results of 100 QMLA instances using the genetic algorithm exploration strategy.}
            \textbf{(c)}, The model space in which \gls{qmla} searches.
            (Left) The total model space contains $2^{18}\approx250,000$ candidate models; 
                normally distributed around $\bar{f}=0.5 \pm 0.14$. 
            (Centre), The models explored during the model search of all instances combined, 
                $\{\hat{H}_i\}$, show that \gls{qmla} tends towards stronger models overall, 
                with models considered having $\bar{f} = 0.76 \pm 0.15$ from $\sim~43,000$ chromosomes across the instances, 
                i.e. each instance trains $\sim~430$ distinct models.                
            (Right), Champion models from each instance, showing \gls{qmla} finds strong models 
                in general, and in particular finds the true model $\ho$ (with $f=1$) in $72\%$ of cases, 
                and $f \geq 0.88$ in all instances. 
            \textbf{(d)}, Hinton diagram showing the rate at which each term is found within the winning model, $\hp$. 
                The size of the blocks show the frequency with which they are found, while the colour indicates 
                whether that term was in the true model (blue) or not (red).
                Terms represent the coupling between two qubits, e.g $\hat{\sigma}_{(1, 3)}^{x}$ 
                    couples the first and third qubits along the $x$-axis. 
                We test four qubits with full connectivity, resulting in 18 unique terms 
                (terms with black rectangles are not considered by the \gls{ga}).
        }
\end{figure*}

\glsreset{ga}


In general, the \emph{model space} grows combinatorially with the dimension of $Q$, 
    and it is not feasible to prescribe a deterministic \gls{es} as above, 
    without an excellent guess of which subspace to entertain, 
    or exploring a large fraction of the large space. 
An adaptive \gls{es} is warranted to explore this model space without relying on prior knowledge:
    in order to approximate $\ho$ we devise a \gls{ga} to perform the search. 
In this study we employ interactive quantum likelihood estimation within the model training subroutine (\gls{qhl}), 
    rendering the procedure applicable only in cases where we have control on $Q$, 
    such as where $Q$ is an untrusted quantum simulator~\cite{Wiebe:2014qhl}.  
\glspl{ga} are combinatorial optimisation techniques,
    applicable to problems where a large number of configurations are possible, 
    where it is infeasible to test every candidate:
    their outcome is not guaranteed to be optimal, but is likely to be near-optimal
   ~\cite{holland1992adaptation}.
Such configurations are called \emph{chromosomes},
    whose suitability can be quantified through calculation of some \emph{\gls{of}}.
\Glspl{of} are not necessarily absolute, 
    i.e. it is not always guaranteed that the optimal chromosome should achieve a 
    certain value of the \gls{of}, but rather the chromosome 
    with highest (lowest) observed value is deemed the strongest candidate.
$N_m$ models are proposed in a single \emph{generation} of the \gls{ga},
    which runs for $N_g$ generations, in clear analogy with the branches of an \gls{et}.
We outline the usual process of \gls{ga}s, along with our choice of 
    selection, cross-over and mutation mechanisms, in \ref{sm:ga}.
\par 

First we define the set of \emph{terms}, $\mathcal{T} = \{ \hat{t}_i\}$, which \emph{may} occur in $\ho$, 
    e.g. by listing the available couplings between lattice sites in our target system, $Q$.
$\hat{t}_i$ are mapped to binary \emph{genes} such that chromosomes 
    -- written as binary strings where each bit is a single gene --  
    can be interpreted as models:
    an example of this mapping is shown in \cref{table:chromosome_example}.
We study a 4-site lattice with arbitrary connectivity under a Heisenberg--XYZ model, 
    omitting the transverse field for simplicity, 
    i.e. any two sites $i,j$ can be coupled by $\sx_i \sx_j, \ \sy_i \sy_j \ $ or $ \ \sz_i \sz_j$.
There are therefore $\left| \mathcal{T}\right| = 3 \times {4 \choose 2} = 18$ binary genes. For brevity we introduce the compact notation 
    $\hat{\sigma}^{ab}_{(i,j)} = \hat{\sigma}^{a}_{i}\hat{\sigma}^{a}_{j} + \hat{\sigma}^{b}_{i}\hat{\sigma}^{b}_{j}$. 
We compose the true model of half the available terms, chosen randomly from $\mathcal{T}$, 
\begin{equation}
    \label{eqn:gen_alg_true_model}
    \ho = \s_{(1, 2)}^{yz}\s_{(1, 3)}^{z}\s_{(1, 4)}^{y}\s_{(2, 3)}^{xy}\s_{(2, 4)}^{x}\s_{(3, 4)}^{xz}.
\end{equation} 
\par 

We can view \gls{qmla} as a binary classification routine, 
    i.e. it attempts to classify whether each $\hat{t}_i \in \termset$ is present in $\termseto$;
    equivalently, whether each available interaction occurs in $Q$. 
It is therefore appropriate to adopt metrics from the machine learning literature 
    regarding classification. 
Our primary figure of merit, against which candidate models are assessed,
    is $F_1$-score: the harmonic mean of
    \emph{precision} (fraction of terms in $\termset_i$ which are in $\termset_0$) and 
    \emph{sensitivity} (fraction of terms in $\termset_0$ which are in $\termset_i$), 
    described in full in \ref{sec:f_score}.
In practice, the $F_1$-score $f \in (0,1)$ indicates the degree to which a model captures the physics of $Q$: 
models with $f=0$ share no terms with $\ho$, while $f=1$ is found uniquely for $\ho$. 
Importantly, the $F_1$-score is a measure of the overlap of terms between the true and candidate models: it is not guaranteed that models of higher $f$ out-perform the predictive ability of those with lower $f$.
However we conjecture that predictive power -- i.e. predicting the true system's dynamics as captured by log likelihoods -- improves with $f$, discussed in \ref{sm:f_score_justification}.

\par

\gls{qmla} should not presume knowledge of $\ho$:
    the model search can not verify the presence of any $\hat{t} \in \termset_0$ definitively,
    although the demonstrations in this work rely on simulated $\ho$, 
    so we can assess the performance of \gls{qmla} with respect to $\fs$. 
Without assuming knowledge of $\ho$ while performing the model search, 
    we do not have a natural objective function.
By first simulating $\ho$, however, 
    we can design objective functions which iteratively improve $f$ on successive generations of the \gls{ga}. 
We perform this analysis in \ref{sm:objective_fncs}, 
    and hence favour an approach inspired by Elo ratings, 
    which is the system used in competitions including chess and football to rate individual competitors, 
    detailed in \ref{sec:elo}~\cite{elo1978rating}.
The \emph{Bayes factor enhanced Elo ratings} give a nonlinear points transfer scheme: 
    \begin{easylist}[itemize]
        & all models on $\mu$, $\{ \hi \}$ are assigned initial ratings, $\{R_i\}$;
        & models on $\mu$ are mapped to vertices of a regular graph with partial connectivity;
        & pairs of models connected by an edge, $(\hi, \hj)$, are compared through \gls{bf}, giving $\bij$;
        & the model indicated as inferior by $\bij$ transfers some of its rating to the superior model:
        the quantity transferred, $\Delta R_{ij}$, reflects 
        && the statistical evidence given by $\bij$;
        && the initial ratings of both models, $\{R_i, R_j\}$.
    \end{easylist}
\par 
The ratings of models on $\mu$ therefore increase or decrease depending on their relative performance,
    shown for an exemplary generation in \cref{fig:gen_alg_analysis}\textbf{a}. 
\par 

We use a \emph{roulette selection} for the design of new candidate models:
    two models are selected from $\mu$ to become \emph{parents} and spawn \emph{offspring}. 
The selection probability for each model $\hi \in \mu$ is proportional to $R_i$ after all comparisons on $\mu$;
    the strongest $\nicefrac{N_m}{3}$ models on $\mu$ are available for selection as parents while 
    evidently weaker models are discarded, see \cref{fig:gen_alg_analysis}\textbf{(a - inset)}.
This procedure is repeated until $N_m$ models are proposed; 
    these do not have to be \emph{new} 
    -- \gls{qmla} may have considered any given model previously -- 
    but all models within a generation should be distinct. 
We show the progression of chromosomes for a \gls{qmla} instance in \cref{fig:gen_alg_analysis}\textbf{(b)}, 
    with $N_g=15, N_m=60$. 
\par 

The \gls{ges} employs \emph{elitism} rules, whereby the two models on $\mu$ with highest $R$
    are automatically considered on $\mu+1$.
If a single model is observed as the highest-fitness candidate for five successive generations, 
    the model search terminates.
Finally, the global champion model is set as the highest-fitness model from the final generation, $\hp=\hat{H}_C^{N_g}$.
\par 

In \cref{fig:gen_alg_analysis}\textbf{(c)} we show three different samplings of the model space: (left) considering a random sample of $10^6$ models from all possible models; (centre) restricting to models entertained for at least one generation by \gls{qmla}; and (right) restricting to those models alone, that were found as global champions. 
The data in $\{\hat{H}_i\}$ and $\{\hp\}$ are from 100 independent instances of \gls{qmla}. 
Typical instances of \gls{qmla} train ${\sim}400$ models before declaring $\hp$, representing $0.16\%$ of the total model space, finding $\hp=\ho$ precisely in $72\%$ of instances. 
Moreover, all instances find $\hp$ with $f \geq 0.88$.
Finally, by considering the Hinton diagram, \cref{fig:gen_alg_analysis}\textbf{(d)},
    we see in combining the instances' results that, while some terms are erroneously included in $\hp$, the identification rate of terms $\hat{t}_i \in \termset_0$  ($\geq95\%$) is significantly higher than $\hat{t}_i \notin \termset_0$ ($\leq10\%$). 
\par 
Taken together, the win rates for the most popular models and the rates with which each term is 
    identified allow for post-hoc determination of a single model with high confidence, 
    corresponding to complete characterisation of $Q$.

\section{Discussion}\label{sec:discussion}

We have shown how \gls{qmla} can serve the purpose of characterising a given black box quantum system $Q$ in several scenarios, identifying the most appropriate (Hamiltonian) model to describe the outcomes of experiments performed on states evolved via $Q$. 
The capabilities of \gls{qmla} were showcased in two different contexts: 
(\emph{i}) the selection of the most suitable model, given a prescribed list of options;
(\emph{ii}) the design of an approximate model, starting from a list of simple primitive terms to combine.
In case (\emph{i}) we tested how \gls{qmla} is highly effective at discriminating between wrong and accurate models, with success rates exceeding 60\% in all circumstances, which can be improved upon given access to  additional computational resources. 
Whereas in (\emph{ii}), designing a model ab-initio for the targeted system, we studied how a genetic strategy, combined with an appropriate objective function, can deliver a model embedding all the correct Hamiltonian terms in $\geq 95\%$ of independent instances, whilst exploring a negligible portion ($< 0.2\%$) of a model space including hundreds of thousands of possible models. 
We emphasise the combinatorial growth in the number of models to entertain  with the size of $Q$: the model space grows exponentially when searching for an optimal model via brute-force, whereas the adaptive search can quickly find an approximate -- if not exact -- model. 

\par

In summary, \gls{qmla} provides the infrastructure which facilitates the formulation of candidate models in attempt to explain observations yielded from the quantum system of interest. 
The learning procedure is enabled by informative experiments -- designed by \gls{qmla} to optimally exploit results from the setup, in an adaptive fashion similar to~\cite{krenn2016automated} -- whose outcomes provide the feedback upon which the agent is trained.
In order to do so, 
\gls{qmla} always assumes access to a same small set of facilities/subroutines: 
(\emph{i}) the ability to prepare $Q$ in a set of fiducial \emph{probe} input states; 
(\emph{ii}) the ability to perform projective measurements on the quantum states evolved by $Q$;
(\emph{iii}) the availability of a trusted simulator 
-- either a classical computer or quantum simulator -- capable of reproducing the same dynamics as $Q$ and
(\emph{iv)} a subroutine for the training of model parameters, e.g. \gls{qhl}. 
We stress how these requirements are readily verified in a wide class of experimental setups.
The only challenging requirement is \emph{(iii)}, due to the well-known inefficiency of classical devices at simulating generic quantum systems. Nevertheless, we argue how the development of quantum technologies might provide reliable quantum simulators, which could then be deployed to characterise batches of equivalent devices.
In this context, we envisage how our protocol might be readily applied in circumstances of crucial interest. 
One such example is offered by the identification of spurious interactions in an experimental quantum device. 
In prototyping many quantum devices, it is likely to observe unwanted terms, e.g. cross-talk among qubits, or artifacts arising from workarounds in devices with limited connectivity~\cite{linke2017experimental, kandala2019error, wright2019benchmarking, ayral2020quantum,vigliar2021error}. 
In this scenario, the user might wish to test whether interactions among $Q$'s components occur which were not designed on-purpose:
this should allow for the calibration or redesign of the device to account for such defects.
\par 

%

One of the most promising applications of \gls{qmla} is the identification of the class of models which most closely represents $Q$'s interactions. 
We envision future work combining the breadth of such a ``family''-search, 
with the depth of the genetic exploration strategy, allowing for rapid classification of black-box quantum systems and devices. 

This would be similar in spirit to recently demonstrated machine-learning of the different phases of matter and their transitions~\cite{carrasquilla2017machine, van2017learning, uvarov2020machine}, and classification of vector vortex beams \cite{giordani2020machine}.
For instance,  considering $Q$ as an untrusted quantum simulator, whose operation the user wishes to verify,  
\gls{qmla} can assess whether the device faithfully dials an encoded $\ho$, and if not, what is the actual $\hp$ implemented. 
Moreover, connected to an online device such as a trusted simulator, \gls{qmla} could monitor whether the device has transitioned to an undesired regime. 
Overall, we expect this functionality to prove helpful beyond the cases shown in this paper, serving as a robust tool in the development of near-term quantum technologies.
 



\clearpage
{\small

\section*{Acknowledgements}
{\small
    The authors thank Stefano Paesani, Chris Granade and Sebastian Knauer  for helpful discussions. 
    BF acknowledges support from Airbus and EPSRC grant code EP/P510427/1.
    This work was carried out using the computational facilities of the Advanced Computing Research Centre, University of Bristol - http://www.bristol.ac.uk/acrc/.
    The authors acknowledge support from the Engineering and Physical Sciences Research Council (EPSRC), Programme Grant EP/L024020/1, from the European project QuCHIP.
    The authors also acknowledge support from the EPSRC Hub in Quantum Computing and Simulation (EP/T001062/1).
    A.L. acknowledges fellowship support from EPSRC (EP/N003470/1).
    } 
    
    \section*{Author Contribution}
    {\small
    BF, AAG, RS and NW conceived the project. 
    BF developed software and performed simulations. 
    AAG, RS and AL supervised the project. 
    All authors contributed to the manuscript.
    }
    \section*{Competing financial interests}
    {\small
    The authors declare no competing financial interests
    }

    \section*{Code Availability}
    {\small
    The source code is available within the \glsentryfull{qmla} framework, an open source Python project~\cite{flynn2021QMLA}.
    }

}

    \clearpage
    \onecolumngrid

    \renewcommand{\thesection}{Appendix~\Alph{section}}
\renewcommand{\thesubsection}{\arabic{subsection}}
\renewcommand*\thefigure{A\arabic{figure}}  
\renewcommand*\thetable{A\arabic{table}}  
\renewcommand{\theequation}{A\arabic{equation}}
\setcounter{section}{0}    
\setcounter{figure}{0}    
\setcounter{equation}{0}
\appendixpageoff
\appendixtitleoff

\glsresetall

\section{Glossary}\label{sm:gloosary}
For reference, here we succinctly define notation as used throughout the text. 

\begin{easylist}[itemize]
    & $Q$ : target quantum system whose model we wish to find.
    & $\hat{H}_i$ : candidate Hamiltonian model.
    & $\hat{H}_0$ : true model, i.e. the Hamiltonian describing the target system, $Q$.
    & $\mathbb{H}$ : set of Hamiltonian models

    & $\termset$ : set of terms, $\{ \hat{t}\}$; refers to the entire space of terms available.
    & $\termseto$ : set of terms in $\ho$.
    & $\termseti$ : set of terms in a candidate model $\hi$. 
    & $\terms_i$ : vector enforcing order on $\termseti$, $\left( \hat{t}_1  \ \dots \ \hat{t}_n \right)$.

    & $\al_i$ : vector of scalar parameters such that $\hi = \al_i \cdot \terms_i$
    & $\al_i^{\prime}$ : optimised parameters for $\hi$ such that $\hi^{\prime} = \al_i^{\prime} \cdot \terms_i$

    & $\al_0$ : parameters of the true model such that $Q$ is described by  $\ho = \al_o \cdot \terms_o$
    & $\al_p$ : parameter vector of a proposal (hypothesis) within \gls{qhl}, i.e. a particle.
    & $P_{\al}$ : prior distribution for the parameterisation of $\hi$.

    & $e$ : single experiment, encompassing all experimental controls such as evolution time and probe state.
    & $\expset$ : set of experiments.
    & $\expset_{i}$ : set of experiments designed specifically during training of a single model $\hi$. 
    & $\expset_{V}$ : set of \emph{validation} expeimrents designed independently (in advance) of each model's training.

    & $d$ : single datum, i.e. measurement of performing $e$ on $Q$.
    & $\expdata$ : set of data, $\{ d \}$, measurements corresponding to some set of experiments $\expset$.

    & $\lk$ : likelihood of measuring a datum $d$ from $Q$ by running an experiment $e$ assuming a hypothesis $\al_p$.
    & $\tll$ : total log likelihood, i.e. the sum of likelihoods with respect to some $\expset$.
    & $\bij$ : Bayes factor between $\hi$ and $\hj$.
\end{easylist}

\section{Genetic Algorithm methods}\label{sm:ga}
Here we describe the \gls{ga} protocol followed, including \gls{qmla}-specific aspects. 
All available Hamiltonian terms compose the set $\termset$, 
    such that models are uniquely identified by chromosomes, 
    written as bit strings where each bit is a \emph{gene} corresponding to a single term.
With $\left|\termset\right|=n$, there are $2^n$ valid chromosomes (bit strings), 
    which constitute the \emph{population} $\population$ of possible models. 
The model search phase is then the standard \gls{ga} procedure:

\begin{easylist}[enumerate]
    & Randomly select $N_m$ chromosomes $\{\hat{H}_i\}$ from $\population$
    && Assign to the first generation: $ \{\hat{H}_i\} \gets \mu$
    & \label{ga:loop} For each $\hi \in \mu$
    && Train $\hi$ via \gls{qhl}
    && Apply objective function: $g(\hi)$
    && Assign selection probability 
    &&& $s_i \sim g(\hi)$. 
    &&& $s_i$ is relative to other models in $\mu$ (i.e. $s_i^{\mu}$ are normalised). 
    & Spawn new models
    && Select two parents from $\mu \rightarrow \hat{p}_A, \hat{p}_B$ 
    &&& according to $s_i$ of each model
    && Cross-over parents to produce children
    &&& $\ttt{C}(\hat{p}_A, \hat{p}_B) \gets \hat{c}_A, \hat{c}_B$
    && Probabilistically mutate children chromosomes
    &&& $\ttt{M}(\hat{c}_A, \hat{c}_B) \gets \hat{c}_A, \hat{c}_B$
    & Collect newly proposed children chromosomes 
    && $\{\hat{c}_i\} \gets \mu + 1$ 
    & Determine top model from this generation
    && $\hat{H}_{C(\mu)} = \argmax\limits_{\hi} \{g(\hi)\}$.
    &
    && \ttt{if} $\hat{H}_{C}^{\mu} == \hat{H}_{C}^{\mu - 4}$ 
    &&& (strongest model not improved upon in five generations)
    &&& move to \ref{ga:finish}
    && \ttt{else if} $\mu == N_g$ 
    &&& (the max allowed generations)
    &&& move to \ref{ga:finish}
    && \ttt{else}
    &&& $\mu \gets \mu + 1$ 
    &&& return to \ref{ga:loop}
    & \label{ga:finish} Nominate the strongest model from the final generation
    && $\hp = \argmax\limits_{\hat{H}_{C}^{\mu}} \{g(\hat{H}_{C}^{\mu})\}$.
\end{easylist}

\subsection{Selection}
Selection of parents from $\mu$ is through roulette selection:
    all models within a generation are assigned selection probabilities,
    $s_i^{\mu}$, reflecting their fitness relative to contemporary models. 
In order to ensure high quality among offspring, 
    we truncate the $N_m$ models in a single generation, 
    such that selection occurs out of only the $N_m^{\prime}$ strongest models, 
    e.g. $N_m^{\prime} = \nicefrac{N_m}{2}$. 
Examples of $s_i$ can be seen in the inset of \cref{fig:gen_alg_analysis}(a), 
    showing the chance of each model becoming a parent, 
    coloured by its $F_1$-score.
Crucially, models are selected in pairs: 
    the probability of the pair  $(\hi, \hj)$ 
    being chosen as parents is proportional to $s_i s_j$.
This is important because in some cases, one (or few) models can dominate, such that 
    $s_i \gg s_j \gg s_k$: in these cases, the probabilities of the inferior models
    still matter, e.g. $ s_i s_j \gg s_j s_k$. 
Further, the same pair of models can reproduce multiple times by crossing over at different 
    positions.
e.g. if the model space uses a chromosome of length 6, $(\hi, \hj)$  can produce offspring
    via one--point crossovers at positions 3 or 4;
    see \cref{table:chromosome_example} for an example of crossover at position 3. 
\par
\subsection{Crossover}    
Parents produce offspring through a \emph{one-point} crossover, 
    i.e. the first $\kappa$ genes from $p_A$ are conjoined with the latter $n - \kappa$ of $p_B$,
    and we choose $\kappa \in \left( \frac{n}{4}, \frac{3n}{4} \right)$ randomly for each pair of parents. 
Given that $\kappa$ is not fixed, the same pair of parents can be selected multiple times with different $\kappa$.
Each gene from each child chromosome has the opportunity to mutate, 
    according to a $25\%$ mutation probability.

\subsection{Objective function}
There is no known natural metric which can capture definitively the suitability of candidate models, 
    although some proxies can be computed.
Intuitive examples that evaluate the candidate's reproduction of the target system's data are the the log--likelihood or $R^2$; 
    these are problematic insofar as they demand specification of input states and evolution times for their calculation, 
    introducing bias. 
Moreover, our model training subroutine results in imperfect paramterisations $\vec{\alpha}_i^{\prime}$, 
    making it challenging to evaluate models in absolute terms.
We outline a number of possible objective functions in  \ref{sm:objective_fncs}, 
    ultimately favouring the Bayes-factor enhanced Elo ratings scheme laid out in \ref{sec:elo}.
\par 

\subsection{Implementation}
The number of models entertained per generation, $N_m$, and the maximum number of generations, $N_g$,
    influence how far the model search can reach:
    increasing $N_m$ allows for a greater diversity of models in the gene pool, 
    while increasing $N_g$ allows precision by considering more models similar to those favoured already. 
Clearly this can lead to a cumbersome number of models to train and evaluate, 
    which must be balanced in order to keep the algorithm tractable. 
To do so, we allow a generous $N_g$ but include an \emph{elite} cutoff:
    if the strongest model is not replaced after 5 generations, the model search terminates. 
For practical reasons we use $N_m = 60$ and $N_g=32$.

\section{Subroutines}\label{sm:subroutines}
\subsection{Quantum Hamiltonian Learning}\label{sm:qhl}
First suggested in \cite{Granade:2012kj} and since developed \cite{wiebe2014qhlpra, Wiebe:2014qhl} 
    and implemented \cite{wang2017experimental,gentile2020learning}, 
\gls{qhl} is a machine learning algorithm for the optimisation of a given Hamiltonian parameterisation 
    against a quantum system whose model is known apriori. 
Given a target quantum system $Q$ known to be described by some Hamiltonian $\hat{H}(\vec{\alpha})$, 
    \gls{qhl} optimises $\vec{\alpha}$.
This is achieved by interrogating $Q$ and comparing its outputs against proposals $\vec{\alpha}_p$. 
In particular, an experiment is designed, consisting of an input state, $\ket{\psi}$, and an evolution time, $t$.
This experiment is performed on $Q$, whereupon its measurement yields the datum $d \in \{0, 1\}$, 
    according to the expectation value $\left| \bra{\psi} e^{-i \ho t} \ket{\psi} \right|^2$. 
Then on a trusted (quantum) simulator, proposed parameters $\vec{\alpha}_p$ are encoded to the 
    known Hamiltonian, and the same probe state is evolved for the chosen $t$ and projected on to $d$, 
    i.e. $\left| \bra{d} e^{-i \hat{H}(\vec{\alpha}_p) t } \ket{\psi} \right|^2 $ is computed.
The task for \gls{qhl} is then to find $\vec{\alpha}^{\prime}$ for which this quantity 
    is close to 1 for all values of $(\ket{\psi}, t)$, 
    i.e. the parameters input to the simulation produce dynamics consistenst with those measured from $Q$.
\par 

The procedure is as follows. 
A \emph{prior} probability distribution $P_{\alpha}$ of dimension $\left| \al \right|$ 
    is initialised to represent the constituent parameters of $\al$. 
$P_{\alpha}$ is typically a multivariate normal (Gaussian) distribution; 
    it is therefore necessary to pre-suppose some mean and width for each parameter in $\al$. 
This imposes prior knowledge on the algorithm whereby the programmer must decide the range in 
    which parameters are \emph{likely} to fit:
    although \gls{qhl} is generally robust and capable of finding parameters outside of this prior,
    the prior must at least capture the order of magnitude of the target parameters. 
An example of imposing such domain-specific prior knowledge is, 
    when choosing the prior for a model representing an $e^-$ spin in a \gls{nv} centre,
    to select $GHz$ parameters for the electron spin's rotation terms, and $MHz$ terms 
    for the spin's coupling to nuclei, as proposed in literature. 
It is important to understand, then, that \gls{qhl} removes the prior knowledge 
    of precisely the parameter representing an interaction in $Q$, but does rely on a ball-park estimate thereof from which to start. 
\par 

In short, \gls{qhl} samples parameter vectors $\al_p$ from $P_{\alpha}$, 
    simulates experiments by computing the \emph{likelihood} $\left| \bra{d} e^{-i \hat{H}(\al_p)t} \ket{\psi} \right|^2$
    for experiments ($\ket{\psi}, t)$ designed by a \gls{qhl} heuristic subroutine, 
    and iteratively improves the probability distribution of the parameterisation $P_{\alpha}$ 
    through standard \emph{Bayesian inference}. 
A given set of $(\ket{\psi}, t)$ is called an experiment, since it corresponds to preparing, evolving and measuring $Q$ once \ \footnote{experimentally, this may involve repeating a measurement many times to determine a majority result and to mitigate noise}. 
\gls{qhl} iterates for $N_E$ experiments. 
The parameter vectors sampled are called \emph{particles}: there are $N_P$ particles used per experiment. 
Each particle used incurs one further calculation of the likelihood function -- 
    this calculation, on a classical computer, is exponential in the number of qubits of the model under consideration
    (because each unitary evolution relies on the exponential of the $2^n \times 2^n$ Hamiltonian matrix of $n$ qubits). 
Likewise, each additional experiment incurs the cost of calculation of $N_P$ particles, 
    so the total cost of running \gls{qhl} for a single model is $\sim N_E N_P$.
It is therefore preferable to use as few particles and experiments as possible, 
    though it is important to include sufficient resources that the parameter estimates have the opporunity to converge. 
Access to a fully operational, trusted quantum simulator admits an exponential 
    speedup by simulating the unitary evolution instead of computing the matrix exponential classically.
\par 

\subsubsection{Bayes Rule}
The mechanism by which $P_{\alpha}$ is updated is \emph{Baye's rule}, \cref{eqn:bayes_rule},

\begin{equation}\label{eqn:bayes_rule}
    \Pr(\vec{\alpha} | D; E) = \frac{\Pr(\expdata| \vec{\alpha}; E) \ \Pr(\vec{\alpha})}{\Pr(\expdata|E)}.
\end{equation}

Bayes rule can be discretised to the level of single particles (individual vectors in the parameter space), sampled from $P_{\al}$
\begin{equation}\label{eqn:particle_likelihood}
    \Pr(\al_p | d; e) = \frac{ \Pr(d | \ \al_p; \ e ) \ \Pr(\al_p) } {\Pr(d | e)}
\end{equation}

where 
\begin{easylist}[itemize]
    & $\al_p$ is the \emph{hypothesis}, i.e. a single parameter vector, called a particle, sampled from $P_{\al}$;
    & $\Pr(\al_p | d; e)$ is the \emph{updated} probability of this particle following the experiment $e$, 
        i.e. accounting for new datum $d$, the probability that $\al=\al_0$;
    & $\Pr(d |\al_p ; e)$ is the likelihood function, 
        i.e how likely it is to have measured the datum $d$ from the system assuming $\al_p$ are the true parameters
        and the experiment $e$ was performed; 
    & $\Pr(\al_p)$ is the probability that $\al_p=\al_0$ according to the prior distribution $P_{\al}$, 
        which we can immediately access; 
    & $\Pr(d|e)$ is a normalisation factor.
\end{easylist}

In order to compute the updated probability for a given particle, then, all that is required is the likelihood function 
    $\Pr(d | \al; e)$. 
This is equivalent to the expectation value of projecting $\ket{\psi}$ onto $d$, after evolving $\hat{H}(\al_p)$ for $t$,
    $\left| \bra{d} e^{-i \hat{H}(\al_p)t} \ket{\psi} \right|^2$, which can be simulated. 
It is necessary first to know the datum $d$ (either 0 or 1) which was projected by $Q$ under real experimental conditions. 
Therefore we first perform the experiment $e$ on $Q$ 
    (preparing the state $\ket{\psi}$ evolving for $t$ and projecting again onto $\bra{\psi}$)
    to retrieve the datum $d$. 
$d$ is then used for the calculation of the likelihood for each particle sampled from $P_{\al}$. 
Each particle's probability can be updated by \cref{eqn:particle_likelihood}, 
    allowing us to redraw the entire probability distribution -- i.e. we compute a \emph{posterior} probability distribution
    by performing this routine on $N_P$ particles.

\subsubsection{Sequential Monte Carlo}

In practice, \gls{qhl} samples from and updates $P_{\al}$  via \gls{smc}.
\gls{smc} samples the $N_P$ particles from $P_{\al}$, and assigns each particle a weight, $w_0 = 1/N_P$.
Each particle corresponds to a unique position in the parameters' space, i.e. $\al_p$.
Following the calculation of $\Pr(\al_p | d; e)$, 
    the weight of particle $p$ are updated by \cref{eqn:qhl_weights}.
\begin{equation}\label{eqn:qhl_weights}
    w_p^{new} = \Pr(\al_p | d; e) \times w_p^{old}
\end{equation}

\begin{figure}
    \centering
    \includegraphics{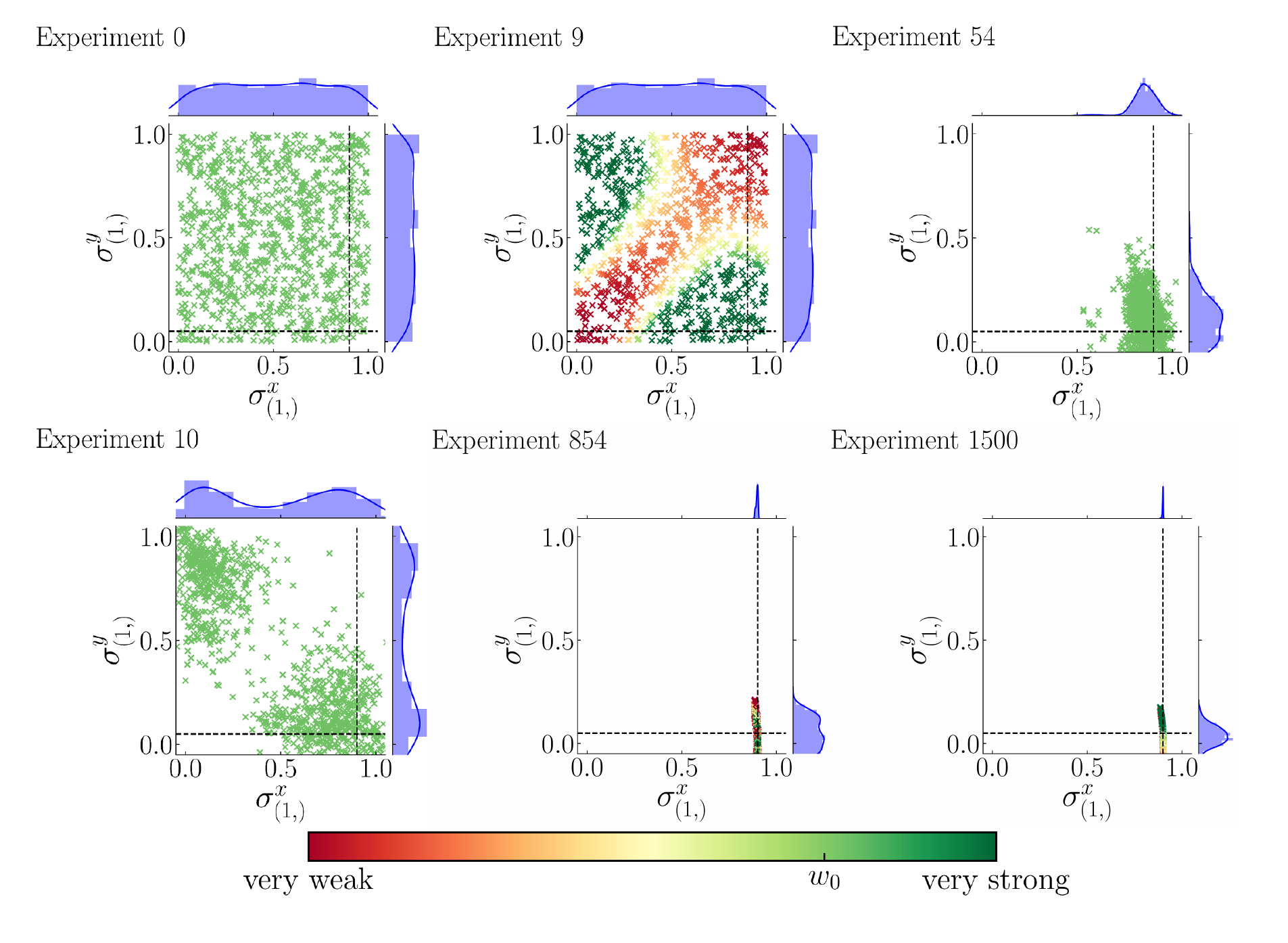}
    \caption{
        \gls{qhl} via \gls{smc}. 
        The studied model has two terms, $\{\sx, \sy\}$ with true parameters $\alpha_{x}=0.9, \alpha_y=0.05$ (dashed lines). 
        Crosses represent particles, while the distribution $P(\alpha_p)$ for each 
            parameter can be seen along the top and right-hand-sides of each subplot. 
        Both parameters are assigned a uniform probability distribution $\mathcal{U}(0,1)$, representing our prior knowledge of the system. 
        \textbf{(a), }\gls{smc} samples $N_P$ particles from the initial joint probability distribution, 
            with particles uniformly spread across the unit square, each assigned the starting \emph{weight} $w_0$. 
            At each experiment $e$, each of these particles' likelihood is computed according to \cref{eqn:particle_likelihood}
            and its weight is updated by \cref{eqn:qhl_weights}.
        \textbf{(b),} after 9 experiments, the weights of the sampled particles are sufficintly informative that we know we can 
            discard some particles while most likely retaining the true parameters. 
        \textbf{(c),} \gls{smc} resamples according the current $P_{\al}$, 
            i.e. having accounted for the experiments and likelihoods observed to date, 
            a new batch of $N_P$ particles are drawn, and each reassigned weight $w_0$, 
            irrespective of their weight prior to resampling.  
        \textbf{(d, e),} Afer further experiments and resamplings, \gls{smc} narrows $P_{\al}$ to a region around the true parameters. 
        \textbf{(f),} The final \emph{posterior} distribution consists of two narrow distributions centred on $\alpha_x$ and $\alpha_y$. 
    }
    \label{fig:qhl_smc}
\end{figure}

In this way, strong particles (high $\Pr(\al_p | d; e)$) have their weight increased, 
    while weak particles (low $\Pr(\al_p | d; e)$) have their weights decreased. 
Within a single experiment, the weights of all $N_P$ particles are updated thus:
    we \emph{simultaneously} update sampled particles' weights as well as $P_{\al}$. 
This iterates for the following experiment, using the \emph{same} particles.
Eventually, the weights of most particles fall below a threshold, 
    meaning that only a fraction (usually half) of particles have reasonable likelihood of being $\al_0$. 
At this stage, \gls{smc} \emph{resamples}, i.e. selects new particles, according to the updated $P_{\al}$.
Then, the new particles are in the range of parameters which is known to be more likely, 
    while particles in the region of low-weight are effectively discarded. 
This procedure is easiest understood through the example presented in \cref{fig:qhl_smc}, 
    where a two-parameter Hamiltonian is learned starting from a uniform distribution. 

\glsreset{qhl}
\glsreset{smc}

\subsubsection{Log--likelihood}\label{sec:log_likelihoods}
A common metric in statistics is that of \emph{likelihood} $\lk$: 
    the measure of how well a model performs against system data. 
Likelihoods are strictly positive, and because the natural logarithm is a monotonically increasing function, 
    it is equivalent to take the \emph{log--likelihood}, 
    since $\ln(\lk_a) > \ln(\lk_b) \iff \lk_a > \lk_b$. 
Log likelihoods are also beneficial in simplifying calculations, 
    and are less susceptible to system underflow, 
    i.e. very small values of $\lk$ will exhaust floating point precision, 
    but $\ln(\lk)$ will not. 
\par 
We used the notion of likelihood earlier to compute $w_p^{new}$ for each particle in the Bayesian inference within \gls{qhl}. 
For each experiment, we use likelihood as a measure of how well the distribution performed
    considering datum $d$ measured from the system, 
    i.e. we care about how well all particles perform as a collective. 
Therefore for an individual experiment $e$, we take the likelihood as the sum of the new weights of all particles, 
    $\{p\}$,
\begin{equation}
    \label{eqn:exp_likelihood}
    \lk_{e} = \sum\limits_{\{p\}} \Pr(\al_p | d; e) \times w_p^{old}.
\end{equation}
Note we know that
\begin{align}    
    \begin{split}
        w_p^0 = \frac{1}{N_P} & \implies \sum_p^{N_P} w_p^0 = 1 ; \\
        \Pr(\al_p | d; e) \leq 1 & \implies \Pr(\al_p | d; e) \times w_p^{old} \leq w_p^{old} \\
        & \implies \sum\limits_{\{p\}} \Pr(\al_p | d; e) \times w_p^{old} \leq \sum\limits_{\{p\}} w_p^{old} \leq \sum_p^{N_P} w_p^0; \\
        & \implies \lk_{e} \leq 1.
    \end{split}
\end{align}

\cref{eqn:exp_likelihood} essentially says, then, that a good batch of particles, 
    where on average particles perform well, 
    will mean that most $w_i$ are high, so $\lk_{e} \approx 1$. 
Conversely, a poor batch of particles will have low average $w_i$, so $\lk_{e} \approx 0$. 
\par

In order to assess the quality of a \emph{model}, $\hi$, 
    we can consider the performance of a set of particles throughout a set of experiments $\expset$, 
    through its \gls{tll},
\begin{equation}
    \label{eqn:log_total_likelihood}
    \tll_{i} = \sum\limits_{e \in \expset} \ln (\lk_{e}).    
\end{equation}

The set of experiments on which $\tll_{i}$ is computed, $\expset$, 
    as well as the particles whose sum constitute each $\lk_{e}$
    can be the same experiments on which $\hi$ is trained, $\expset_i$, but in general need not be, 
    i.e. $\hi$ can be evaluated by considering different experiments than those on which it was trained. .
For example, $\hi$ can be trained with $\expset_i$ to optimise $\al_i^{\prime}$, 
    and thereafter be evaluated using a different set of experiments $\expset_v$, 
    such that $\tll_{i}$ is computed using particles sampled from the distribution after optimising $\al$, 
    $\Pr(\al_i^{\prime})$, and may use a different number of particles than the training phase. 

Perfect agreement between the model and the system would result in $\lk_{e}=1 \Rightarrow \ln(\lk_{e})=0$, 
    as opposed to poor agreement $\lk_{e} < 1 \Rightarrow \ln(\lk_{e}) < 0$.
Then, in all cases \cref{eqn:log_total_likelihood} is negative, 
    and across a series of experiments,
    strong agreement gives low $\left| \tll_{i} \right| $, 
    whereas weak agreement gives large $\left| \tll_{i} \right| $.

\subsection{Model selection and Bayes factors}\label{sm:bayes_factors}
Finally, we can use the above tools to \emph{compare} models. 
It is statistically meaningful to compare models via log--likelihoods 
    if and only if they have considered the same data, 
    i.e. if models have each attempted to account for some experiments 
    $\expset$.
Given a set of models which have all been evaluated against some $\expset_v$, 
    it is straightforward to find the strongest model by choosing the model with the highest \gls{tll}. 
Of course it is first necessary to ensure that each model has  
    been adequately trained: while inaccurate models are unlikely to strongly 
    capture the system dynamics, they should first train on the system 
    to determine their best attempt at doing so, 
    i.e. they should undergo, until convergence, a parameter learning process (as in \cref{fig:qhl_smc}).
\par 

We can also exploit direct pairwise comparisons between models, however, 
    which need not be the pre-computed $\expset_v$ introduced already;
    we merely must impose that both models' log likelihoods are computed based 
    on any set of experiments $\expset$, with corresponding measurements $\expdata = \{ d_{e}\}_{e \in \expset}$.
Pairwise comparisons can be quantified by the \gls{bf},
\begin{equation}
    \label{eqn:bayes_factors}
    B_{ij} = \frac{\Pr(\expdata | \hi; \expset)}{\Pr(\expdata | \hj; \expset)}.
\end{equation}

We have that, for independent experiments, 
\begin{align}
    \begin{split}
        \Pr(\expdata | \hi; \expset) &= \Pr(d_n | \hi; e_n) \times \Pr(d_{n-1} | \hi; e_{n-1}) \times \dots \times \Pr(d_0 | \hi; e_0) \\
        &= \prod_{e \in \expset} \Pr(d_e | \hi; e)
        = \prod_{e \in \expset} (\lk_{e})_i.
    \end{split}
\end{align}

We also have, from \cref{eqn:log_total_likelihood}
\begin{align}
    \label{eqn:log_likelihood}
    \begin{split}
        \tll_{i} &= \sum_{e \in \expset} \ln\left(\left(\lk_{e}\right)_i\right) \\
        \implies e^{\tll_{i}} = \exp \left(\sum_{e \in \expset} \ln\left[\left(\lk_{e}\right)_i\right]\right)
        &= \prod_{e \in \expset} \exp \left( \ln \left[ (\lk_{e})_i \right]  \right)
        = \prod_{e \in \expset} (\lk_{e})_i.
    \end{split}
\end{align}

So we can write 
\begin{align}
    \label{eqn:bf_by_ll}
    \begin{split}
        \bij &= \frac{\Pr(\expdata | \hi; \expset)}{\Pr(\expdata | \hj; \expset)} 
        = \frac{ \prod_{e \in \expset} (\lk_{e})_i } {\prod_{e \in \expset} (\lk_{e})_j }
        = \frac{e^{\tll_{i}}}{e^{\tll_{j}}}
    \end{split}
\end{align}

\begin{equation}
    \label{eqn:bf_succinct}
    \implies \bij = e^{\tll_{i} - \tll_{j}}    
\end{equation}

This is simply the exponential of the difference between two models' total log-likelihoods when presented the same set of experiments. 
Intuitively, if $\hi$ performs well, and therefore has a high \gls{tll}, $\tll_{i}=-10$, 
    and $\hj$ performs worse with $\tll_{j}=-100$, then $B_{ij} = e^{-10-(-100)} = e^{90} \gg 1$.
Conversely for $\tll_{i}=-100, \tll_{j}=-10$, then $B_{ij} = e^{-90} \ll 1$. 
Therefore $\left| B_{ij} \right|$ quantifies how strong the statistical evidence is in favour (or against) a model $\hi$ compared to an alternative hypothetical $\hj$:

\begin{equation}
    \label{eqn:bf_cases}
    \begin{cases}
        B_{ij} > 1 & \Rightarrow \hi \text{\ stronger than \ } \hj, \\
        B_{ij} < 1 & \Rightarrow \hj \text{\ stronger than \ } \hi, \\
        B_{ij} = 1 & \Rightarrow \hi \text{\ as strong as \ } \hj, \\
    \end{cases}
\end{equation} 

\par 
Finally, as mentioned it is necessary for the \gls{tll} of both models in a \gls{bf} calculation to
    refer to the same $\expset$. 
There are a number of ways to achieve this, 
    which we briefly summarise here for reference when discussing objective functions. 
During training (the \gls{qhl} stage described above), candidate model $\hi$ is trained against $\expset_i$, 
    designed by a heuristic to optimise parameter learning specifically for $\hi$;
    likewise $\hj$ is trained on $\expset_j$. 
The simplest method to compute the \gls{bf} is to enforce $\expset=\expset_i \cup \expset_j$ 
    in \cref{eqn:bayes_factors}, i.e. to cross-train $\hi$ using the data designed specifically for training $\hj$, 
    and vice versa. 
This is a valid approach because it challenges each model to attempt to explain experiments
    designed explicitly for its competitor,   
    at which only truly accurate models are likely to succeed. 
\par 
A second approach builds on the first, but incorporates \emph{burn--in} time in the training regime:
    this is a standard technique in the evaluation of \gls{ml} models whereby its earliest iterations 
    are discounted for evaluation so as not to skew its metrics, 
    ensuring the evaluation reflects the strength of the model. 
In \gls{bf}, we achieve this by basing the \gls{tll} only on a subset of the training experiments. 
For example, the latter half of experiments designed during the training of $\hi$, $\expset_i^{\prime}$. 
This does not result in less predictive \gls{bf}, since we are merely removing the 
    noisy segments of the training for each model. 
Moreover it provides a benefit in reducing the computation requirements: 
    updating each model to ensure the \gls{tll} is based on $\expset^{\prime}=\expset_i^{\prime} \cup \expset_j^{\prime}$
    requires only half the computation time, 
    which can be further reduced by lowering the number of particles used during the update, $N_P^{\prime}$, 
    which will give a similar result as using $N_P$, assuming the posterior has converged.
\par 

A final option for the set of experiments is to design a set of experiments, $\expset_v$, 
    that are valid for a broad variety of models, and so will not favour any particular model.
This is clearly a favourable approach: 
    provided for each model we compute \cref{eqn:log_total_likelihood} using $\expset_v$,
    we can automatically select the strongest model based solely on their \gls{tll}s, 
    meaning we do not have to perform further computationally-expensive updates, 
    as required to cross-train on opponents' 
    experiments during \gls{bf} calculation. 
However, it does impose on the user to design a \emph{fair} $\expset_v$, 
    requiring unbiased probe states $\{\ket{\psi}\}$ and times $\{t\}$ on a timescale 
    which is meaningful to the system under consideration. 
For example, experiments with $t > T_2$, the decoherence time of the system, 
    would result in measurements which offer little information, 
    and hence it would be difficult to extract evidence in favour of any 
    model from experiments in this domain.
It is difficult to know, or even estimate, such meaningful time scales a priori,
    so it is difficult for a user to design $\expset_v$. 
Additionally, the training regime each model undergoes during \gls{qhl}
    is designed to provide adaptive experiments that take into account
    the specific model entertained, to choose an optimal set of evolution times.
In the case where such constraints can be accounted for, however, this approach is favoured,
    e.g. an experiment repeated in a laboratory where the available 
    probe states are limited and the timescale achievable is understood.

    \section{Exploration Strategies}\label{sm:exploration_strategies}
An \gls{es} specifies the rules by which the model search proceeds for on \gls{et} within \gls{qmla}. 
It can be thought of as a separate utility which \gls{qmla} calls upon for 
    algorithmic parameters and subroutines.   
Within this context, the \gls{es} has several key responsibilities, which we detail below, summarised as:
\begin{easylist}[enumerate]
    \ListProperties(Numbers=r)
    & model generation: 
        combining the knowledge progressively acquired by \gls{qmla} into constructing new candidate models;
    & decision criteria for the model search phase:
        instructions for how \gls{qmla} should respond at predefined junctions, 
        e.g. whether to cease the model search after a branch has completed;
    & true model specification:
        detailing the terms and parameters which constitute $\ho$, in the case where $Q$ is simulated;
    & modular functionality: 
        subroutines called throughout \gls{qmla} are interchangeable such that each \gls{es} specifies the 
        set of functions to achieve its goals.
\end{easylist}

\subsection{Model generation}
The most important function for any \gls{es} is to design new candidate models to entertain in the \gls{qmla} search.
Model generation can be adaptive, i.e. utilise information gathered so far by \gls{qmla}, 
    such as the outcomes of previous models' training, 
    and the results of comparisons between models to date. 
Conversely, model generation could also be non-adaptive: either because following a fixed schedule determined in advance, or be entirely random.
This alludes to the central design choice in composing an \gls{es}: 
how broad should the searchable \emph{model space} be, considering that adequately training each model
is expensive, and that model comparisons are similarly expensive. A thorough estimate is provided in \ref{sec:computationcost}.

Therefore, it represents a crucial design decision, whether timing considerations are more important than thoroughly exploring the model space: huge computational advantages arise from performing but a small subset of all possible model comparisons. 
The approach in \ref{sec:computationcost}~\ref{gr:predefined} is clearly limited in its applicability, 
    mainly in that there is a heavy requirement for prior knowledge, 
    and it is only useful in cases where we either know $\ho \in \mathbb{H}$, 
    or would be satisfied with approximating $\ho$ as the closest available $\hj \in \mathbb{H}$. 
On the opposite end of this spectrum, \ref{sec:computationcost}~\ref{gr:generative_sparse}  is an excellent approach
    with respect to minimising prior knowledge required by the algorithm, 
    although at the significant expense of testing a much larger number of candidate models. 
There is no optimal strategy for all use--cases: 
    specific quantum systems of study demand particular considerations, 
    and the amount of prior information available informs how wide the model search should reach. 

\par 

In this work we have used two straightforward model generation routines.
Firstly, during the study of various physical classes (Ising, Heisenberg, Fermi-Hubbard), 
    a list of lattice configurations were chosen in advance,
    which were then mapped to Hamiltonian models.
This \gls{es} is non-adaptive and indeed the model search
    consists merely of a single branch with no subsequent calls to the model generation routine, 
    as in \ref{sec:computationcost}~\ref{gr:predefined}. 
In the latter section instead we use a \gls{ga}:
    this is clearly a far more general strategy, at a significant computational cost, 
    but is suitable for systems where we have less knowledge in advance. 
In this case, new models are designed based heavily on results from earlier branches of the \gls{et}. 
The genetic algorithm model generation subroutine is listed in \cref{alg:generate_models}, 
    and can broadly be summed up thus: 
    the best models in a generation $\mu$ produce offspring which constitute models on the next generation.
These types of evolutionary algorithms ensure that newly proposed candidates inherit 
    some of the structure which rendered previous candidates (relatively) successful, 
    in the expectation that this will yield ever-stronger candidates.

\subsection{Decision criteria for the model search phase}
Further control parameters, which direct the growth of the \gls{et}, are set within the \gls{es}.
At several junctions within \cref{alg:qmla}, \cref{alg:model_search}, 
    \gls{qmla} queries the \gls{es} in order to decide what happens next.
Here we list the important cases of this behaviour. 

\begin{easylist}[enumerate]
    \ListProperties(Numbers=r)
    & Parameter-learning settings
    && such as the prior distribution to assign each parameter during \gls{qhl}, and the parameters needed to run \gls{smc}.
    && the time scale on which to examine $Q$.
    && the input probes to train upon. 
    & Branch comparison strategy
    && How to compare models within a branch (or \gls{qmla} layer). 
        All methods in \ref{sm:objective_fncs} can be thought of as branch comparison strategies. 
    & Termination criteria
    && e.g. instruction to stop after a fixed number of iterations, or when a certain fitness has been reached.
    & Champion nomination
    && when a single tree is explored, identify a single champion from the branch champions
    && if multiple trees are explored, how to compare champions across trees. 
\end{easylist}

\subsection{True model specification}
It is necessary also to specify details about the true model $\ho$, 
    at least in the case where \gls{qmla} acts on simulated data. 
Within the \gls{es}, we can set $\terms_0$ as well as $\al_0$. 
For example, where the target system is an untrusted quantum simulator $S_u$ to be characterised by interfacing with a trusted (quantum) simulator $S_t$, we decide some $\ho$ in advance. The model training subroutine then calls for likelihoods (see \ref{sm:qhl}): those corresponding to $\ho$ are computed via $S_u$, whereas particles' likelihoods are estimated with $S_t$. 

\subsection{Modular functionality}
Finally, there are a number of fundamenetal subroutines which are called upon throughout the \gls{qmla} algorithm. 
These are written independently such that each subroutine has a number of available implementations. 
These can be chosen to match the requirements of the user, and are set via the \gls{es}. 

\begin{easylist}[enumerate]
    \ListProperties(Numbers=r, Numbers2=l,)
    & Model training procedure
    && i.e. whether to use \gls{qhl} or quantum process tomography, etc. 
    && In this work we always used \gls{qhl}. 
    & Likelihood estimation,
        which ultimately depends on the measurement scheme. 
    && By default, here we use projective measurement back onto the input probe state, 
        $\left| \bra{\psi} e^{-i\hat{H}t} \ket{\psi} \right|^2$.
    && It is possible to change this to implement any measurement procedure, 
        for example an experimental procedure where the environment is traced out. 
    & \label{item:probes} Probes: defining the input probes to be used during training. 
        && In general it is preferable to use numerous probes in order to avoid biasing particular terms. 
        && In some cases we are restricted to a small number available input probes, e.g. to match experimental constraints.
    & Experiment design heuristic: bespoke experiments to maximise the information 
        on which models are individually trained.
        && In particular, in this work the experimental controls consist solely of $\{ \ket{\psi}, t \}$. 
        && Currently, probes are generated once according to \ref{item:probes}, 
            but in principle it is feasible to choose optimal probes based on available or hypothetical information. 
            For example, probes can be chosen as a normalised sum of the candidate model's eigenvectors.
        && Choice of $t$ has a large effect on how well the model can train. 
            By default times are chosen proportional to the inverse of the 
            current uncertainty in $\al$ to maximise Fischer information, 
            through the multi-particle guess heuristic \cite{Wiebe:2014qhl}.
            Alternatively, times may be chosen from a fixed set to force \gls{qhl} to 
            reproduce the dynamics within those times' scale. 
            For instance, if a small amount of experimental data is available offline, 
            a na{\"i}ve option is to train all candidate models against the entire dataset.  
    & Model training prior: change the prior distribution, e.g. \cref{fig:qhl_smc}(a)
\end{easylist}

\section{Considerations about computational cost}
\label{sec:computationcost}

Let us assume the following conditions:

\begin{easylist}[itemize]
    & all models considered are represented by 4-qubit models;
    && so time evolution requires exponentiating the Hamiltonian matrix of dimension $2^4 \times 2^4$. 
    & each model undergoes a reasonable training regime;
    && $N_E=1000, N_P=3000$ (particles and experiments as defined in \ref{sm:qhl});
    && recall traning relies on computing matrix exponentials $e^{-i\hat{H}t}$ for each particle at each experiment;
    &&& $t_{\textrm{exp}} \sim 10^{-3} s$ per matrix exponential
    && $\implies t_{\textrm{QHL}} = N_E \times N_P \times  t_{exp} = 3000s \sim 1 h $;
    & Bayes factor calculations use 
    && $N_E=500, N_P=3000 $
    && $\implies t_{\textrm{BF}} \sim 2 \times  500 \times 3000 \times  10^{-3} \sim 1 h$;
    & there are 12 available terms
    && allowing any combination of terms, this admits a model space of size $2^{12} = 4096$
    & access to 16 computer cores to parallelise calculations over
    && i.e. we can train 16 models or perform 16 \gls{bf} comparisons in $1h$.
\end{easylist}
\par 

\noindent Then, consider the following model generation/comparison strategies.
\begin{easylist}[enumerate]
    \ListProperties(Numbers1=l, Numbers2=r)
    & \label{gr:predefined} Predefined set of $16$ models, comparing every pair of models
    && Training takes $1h$, and ${16 \choose 2} = 120$ comparisons need $8h$
    && total time is $9h$. 
    & \label{gr:generative_full} Generative procedure for model design, comparing every pair of models,
        running for 12 branches
    && One branch takes $9h \implies$ total time is $12 \times 9 = 108h$; 
    && total number of models considered is $16 \times 12 = 192$. 
    & \label{gr:generative_sparse} Generative procedure for model design, where less model comparisons are needed 
        (say one third of all model pairs are compared),
        running for 12 branches
    && Training time is still $1h$
    && One third of comparisons, i.e. $40$ \gls{bf} to compute, requires $3h$
    && One branch takes $4 h \implies$ total time is $36 h$
    && total number of models considered is also $192$. 
\end{easylist}

    \section{Objective Functions}\label{sm:objective_fncs}
As outlined in the main text, in order to incorporate a \gls{ga} within \gls{qmla}, 
    we first need to design a meaningful \gls{of}. 
Here we present a number of \gls{of}s, 
    and select the most suitable for use in the analysis presented in the main text.
We introduce $\fs$ as a figure of merit for individual models compared against $\ho$;
    here we will rely on $\fs$ to indicate the quality of models produced by each \gls{of}. 
Throughout this section it will be useful to refer to examples to demonstrate the efficacy of each \gls{of}. 
We present a set of exemplary models, along with their fitnesses according to all the 
    \gls{of}s discussed, in \cref{table:objective_functions}; the models are
\begin{equation}
    \label{eqn:obj_fnc_eg_models}
    \begin{split}
        \hat{H}_0 \ &= \ \s_{(1, 2)}^{z}\s_{(1, 3)}^{z}\s_{(2, 3)}^{z}\s_{(2, 5)}^{z}\s_{(3, 5)}^{z};\\
        \hat{H}_a \ &= \ \s_{(1, 5)}^{z}\s_{(3, 4)}^{z}\s_{(4, 5)}^{z}; \\
        \hat{H}_b \ &= \ \s_{(1, 4)}^{z}\s_{(1, 5)}^{z}\s_{(2, 5)}^{z}\s_{(3, 4)}^{z}; \\
        \hat{H}_c \ &= \ \s_{(1, 2)}^{z}\s_{(1, 5)}^{z}\s_{(2, 4)}^{z}\s_{(2, 5)}^{z}\s_{(4, 5)}^{z}; \\
        \hat{H}_d \ &= \ \s_{(1, 3)}^{z}\s_{(1, 4)}^{z}\s_{(1, 5)}^{z}\s_{(2, 4)}^{z}\s_{(2, 5)}^{z}\s_{(3, 4)}^{z}\s_{(3, 5)}^{z}; \\
        \hat{H}_e \ &= \ \s_{(1, 2)}^{z}\s_{(1, 3)}^{z}\s_{(1, 5)}^{z}\s_{(2, 3)}^{z}\s_{(2, 5)}^{z}\s_{(4, 5)}^{z}; \\
        \hat{H}_f \ &= \ \s_{(1, 2)}^{z}\s_{(1, 3)}^{z}\s_{(2, 3)}^{z}\s_{(2, 4)}^{z}\s_{(2, 5)}^{z}\s_{(3, 4)}^{z}\s_{(3, 5)}^{z}. \\
    \end{split}
\end{equation}

\par 

Every \gls{of} computes the \emph{fitness} $g_i$ for each candidate $\hi$.
Because we need to map model fitness to roulette selection probability, 
    we require $g_i>0$ and stronger models should record higher $g_i$.
It is not necessary to bound $0\leq g_i \leq 1$, since probalities will be normalised during selection anyway, 
    although in most cases it is straightforward to restrict to this range for ease of interpretation; 
    we also present the selection probability (\%) that would be suggested by each \gls{of} in \cref{table:objective_functions}.
We also track the cardinality of each models' parameterisations, $k$, i.e. the number of terms in the model.

\subsection{$F_1$-score}\label{sec:f_score}

First we should introduce our primary figure of merit
    which we will use to assess model quality objectively: $\fs$. 
This also allows us to judge the outputs of each \gls{of}:
    those which result in models of higher $\fs$ are clearly favourable.
We emphasise that the goal of this work is to identify the \emph{model} which best describes 
    quantum systems, and not to improve on parameter-learning when given access to particular models, 
    since those already exist to a high standard \cite{wiebe2014qhlpra,bairey2019learning}. 
Therefore we can consider \gls{qmla} as a classification algorithm, 
    with the goal of classifying whether individual terms $\hat{t}$ from a set of available 
    terms $\termset = \{\hat{t}\}$ are helpful in describing data which is generated by $\ho$, 
    which has $\termset_0$. 
Candidate models $\hi$ then have $\termset_i$.
We can assess $\hi$ using standard metrics in the \gls{ml} literature, 
    which simply count the number of terms identified correctly and incorrectly,
    \begin{easylist}[itemize]
        & \gls{tp}: number of terms in $\To$ which are in $\Ti$;
        & \gls{tn}: number of terms not in $\To$ which are also not in $\Ti$;
        & \gls{fp}: number of terms in $\Ti$ which are not in $\To$;
        & \gls{fn}: number of terms in $\To$ which are not in $\Ti$.
    \end{easylist}
\par 
\noindent These quantities allow us to define 
\begin{easylist}
    & \emph{precision}: how precisely does $\hi$ capture $\ho$,
    i.e. if a term is included in $\Ti$ how likely it is to actually be in $\To$, Eqn \ref{eqn:precision};
    & \emph{sensitivity}: how sensitive is $\hi$ to $\ho$, 
    i.e. if a term is in $\To$, how likely $\Ti$ is to include it, \cref{eqn:sensitivity}.
\end{easylist}

\begin{subequations}
    \begin{equation}
        \label{eqn:precision}
        \text{precision} = \frac{\rm{TP}}{\rm{TP} + \rm{FP}}
    \end{equation}
       
    \begin{equation}
        \label{eqn:sensitivity}
        \text{sensitivity} = \frac{\rm{TP}}{\rm{TP} + \rm{FN}}
    \end{equation}
\end{subequations}

Informally, precision prioritises that predicted terms are correct, 
    while sensitivity prioritises that true terms are identified. 
In practice, it is important to balance these considerations. 
$F_{\beta}$-score is a measure which balances these, with $F_1$-score in particular giving them equal importance. 
\begin{equation}
    \label{eqn:f1_score}
    F_1 = \frac{2\times (\rm{precision})\times(\rm{sensitivity})}{(\rm{precision} + \rm{sensitivity})} = \frac{\rm{TP}}{\rm{TP} + \frac{1}{2}(\rm{FP} + \rm{FN})}.
\end{equation}
We adopt \fs as an indication of model quality because we are concerned both with precision and sensitivity, 
    which we can use in the analysis of \gls{of}s. 
Additionally, we can use \fs as a straightforward test of our \gls{ga} within \gls{qmla}: 
    by using \fs \ alone as the \gls{of}, i.e.  no learning/comparison of models, we can show that 
    \gls{qmla} finds the true model even in extremely large model spaces.
Of course in realistic cases we can not assume knoweldge of $\To$ and therefore cannot compute 
    $\fs$, so our search for an effective \gls{of} can be reduced to finding the 
    \gls{of} which correlates most strongly with \fs \ in test-cases.

\subsection{Inverse Log-likelihood}\label{sec:inverse_ll}
We have already introduced the total log-likelihood, $\tll_{i}$, in \ref{sec:log_likelihoods} and suggested that model selection is straightforward 
    provided each candidate model computes a meaningful \gls{tll} based on an evaluation experiment set $\expset_v$.
\gls{tll} are negative and the strongest model has lowest $\left| \tll_{i} \right|$ (or highest $\tll_{i}$ overall),
    so the corresponding \gls{of} for candidate $\hi$ is 
\begin{equation}
    \label{eqn:obj_log_likelihood}
    g_i^{L} = \frac{-1}{\tll_{i}}.
\end{equation}

In our tests, \cref{eqn:obj_log_likelihood} is found to be too generous to poor models, 
    assigning them non-negligible probability. 
Its primary flaw, however, is its reliance on $\expset_v$: 
    in order that the \gls{tll} is significant, it must be based on meaningful experiments, 
    the design of which can not be guaranteed in advance, or at least introduces strong bias.

\renewcommand{\arraystretch}{1.25} 
\setlength{\tabcolsep}{3pt}

\begin{table}
    \begin{center}
    \begin{tabular}{|cc|c|c|c|c|c|c|}
\hline
          &      &                              $\hat{H}_a$ &                              $\hat{H}_b$ &                              $\hat{H}_c$ &                              $\hat{H}_d$ &                              $\hat{H}_e$ &                              $\hat{H}_f$ \\
Method &  &                                          &                                          &                                          &                                          &                                          &                                          \\

          & $F_1$ &                                    $0.0$ &                                    $0.2$ &                                    $0.4$ &                                    $0.5$ &                                    $0.7$ &                                    $0.8$ \\
          & $k$ &                                        3 &                                        4 &                                        5 &                                        7 &                                        6 &                                        7 \\
          & $\overline{l_e}$ &                          $0.86 \pm 0.29$ &                          $0.84 \pm 0.29$ &                          $0.77 \pm 0.27$ &                          $0.78 \pm 0.29$ &                          $0.79 \pm 0.26$ &                          $0.79 \pm 0.26$ \\
          & $\tll_i$ &                                     -143 &                                     -152 &                                     -131 &                                     -150 &                                     -125 &                                     -124 \\
\hline \multirow{2}{*}{Inverse log-likelihood} & $g_i^L$ &                                  0.00698 &                                  0.00659 &                                  0.00766 &                                  0.00669 &                                  0.00803 &                                  0.00804 \\
          & $\%$ &                                       23 &                                        0 &                                       25 &                                        0 &                                       26 &                                       26 \\
\cline{1-8}
\hline \multirow{5}{*}{Akaike Info Criterion} & AIC &                                      293 &                                      311 &                                      271 &                                      313 &                                      261 &                                      263 \\
          & AICc &                                      293 &                                      312 &                                      272 &                                      314 &                                      262 &                                      264 \\
          & $w_i^A$ &                                 1.81e-07 &                                  1.4e-11 &                                  0.00724 &                                 4.15e-12 &                                        1 &                                    0.334 \\
          & $g_i^A$ &                                 1.17e-05 &                                 1.03e-05 &                                 1.35e-05 &                                 1.01e-05 &                                 1.46e-05 &                                 1.43e-05 \\
          & $\%$ &                                       22 &                                        0 &                                       25 &                                        0 &                                       27 &                                       26 \\
\cline{1-8}
\hline \multirow{4}{*}{Bayesian Info Criterion} & BIC &                                      301 &                                      322 &                                      284 &                                      331 &                                      277 &                                      281 \\
          & $w_i^B$ &                                 5.49e-66 &                                 1.26e-70 &                                 1.97e-62 &                                 1.11e-72 &                                 8.43e-61 &                                 8.95e-62 \\
          & $g_i^B$ &                                 1.11e-05 &                                 9.65e-06 &                                 1.24e-05 &                                 9.11e-06 &                                 1.31e-05 &                                 1.27e-05 \\
          & $\%$ &                                       23 &                                        0 &                                       25 &                                        0 &                                       27 &                                       26 \\
\cline{1-8}
\hline \multirow{2}{*}{Bayes factor points} & $g_i^p$ &                                        0 &                                        2 &                                        3 &                                        2 &                                        3 &                                        5 \\
          & $\%$ &                                        0 &                                       13 &                                       20 &                                       13 &                                       20 &                                       33 \\
\cline{1-8}
\hline \multirow{3}{*}{Ranking points} & Ranking &                                        6 &                                        4 &                                        3 &                                        5 &                                        2 &                                        1 \\
          & $g_i^R$ &                                        0 &                                      0.1 &                                      0.2 &                                        0 &                                      0.3 &                                      0.4 \\
          & $\%$ &                                        0 &                                       10 &                                       20 &                                        0 &                                       30 &                                       40 \\
\cline{1-8}
\hline \multirow{3}{*}{Elo rating} & Rating &                                      909 &                                      944 &                                     1042 &                                     1007 &                                     1011 &                                     1084 \\
          & $g_i^E$ &                                        0 &                                       35 &                                      133 &                                       98 &                                      102 &                                      175 \\
          & $\%$ &                                        0 &                                        0 &                                       26 &                                       19 &                                       20 &                                       34 \\
\cline{1-8}
\hline \multirow{3}{*}{Residuals} & mean$\{\tilde{r_p^e}\}$ &                                    0.132 &                                    0.146 &                                    0.114 &                                    0.138 &                                   0.0858 &                                   0.0715 \\
          & $g_i^r$ &                                    0.753 &                                    0.729 &                                    0.785 &                                    0.743 &                                    0.836 &                                    0.862 \\
          & $\%$ &                                       23 &                                        0 &                                       24 &                                        0 &                                       26 &                                       27 \\
\hline
\end{tabular}

    \caption{
        Examples of how each objective function, $g$ as described in \ref{sec:inverse_ll} to \ref{sec:elo},
            assign selection probability (denoted \%) to the same set of candidate models, $\{\hi\}$. 
        For each model we first summarise its
            average likelihood $\lk_{e}$ (\cref{eqn:log_likelihood}),     
            total log-likelihood $\tll_{i}$ (\cref{eqn:log_total_likelihood}), 
            and number of terms $k$.
        We use $n=100$ samples, i.e. $\tll_{i}$ is a sum of $n$ likelihoods.  
    }
    \label{table:objective_functions}
    \end{center}
\end{table}

\subsection{Akaike Information Criterion}\label{sec:akaike_info_criterion}
A common metric in model selection is \gls{aic} \cite{dr2002model}.
Incorporating \gls{ll}, 
    \gls{aic} also objectively quantifies how well a given model accounts for data from the target system,
    and punishes models which use extraneous parameters, 
    by incurring a penalty on the number of parameters, $k$. 
\gls{aic} is given by 
\begin{equation}
    \label{eqn:aic}
    AIC_i = 2k_i - 2 \tll_{i},
\end{equation}

In practice we use a slightly modified form of \cref{eqn:aic} which
    corrects for the number of samples $n=\left|\expset_i\right|$, 
    called the \gls{aicc}, 
\begin{equation}
    \label{eqn:aicc}
    AICC_i = AIC_i + 2k\frac{k+1}{n-k-1}. 
\end{equation}

Model selection from a set of candidates occurs simply by selecting the model with $AICC_{\textrm{min}}$.
A suggestion to retrieve selection probability,
    by using \cref{eqn:aicc} as a measure of \emph{relative likelihood},
    is to compute \emph{Akaike weights} (as defined in in Chapter 2 of \cite{dr2002model}),
\begin{equation}
    \label{eqn:akaike_weights}
    w_i^A = exp \left( \frac{{AICC_{\textrm{min}} - AICC_i}}{2} \right), 
\end{equation}
where $AICC_{\textrm{min}}$ is the lowest $AICC$ observed among the models under consideration
    e.g. all models in a given generation. 

Clearly, Akaike weights impose quite strong penalties 
    on models which do not explain the data well, 
    but also punish models with extra parameters, i.e. overfitting, 
    effectively searching for the strongest and simplest model simultaneously.
The level of punishment for poorly performing models is likely too drastic: 
    very few models will be in a range sufficiently close to $AICC_{\textrm{min}}$ 
    to receive a meaningful Akaike weight, 
    suppressing diversity in the model population.
    Indeed, we can see from \cref{table:objective_functions} that this results in most 
    models being assigned negligible weight, which is not useful for parent selection. 
Instead we compute a straightforward quantity as the \gls{aic}--inspired fitness, \cref{eqn:akaike_fitness},
\begin{equation}
    \label{eqn:akaike_fitness}
    g_i^{A} = \left(\frac{1}{AICc_i}\right)^2,
\end{equation}
    where we square the inverse \gls{aic} to amplify the difference in quality between models, 
    such that stronger models are generously rewarded.

\par 

\subsection{Bayesian Information Criterion}\label{sec:bayes_info_criterion}
Related to the idea of \gls{aic}, \cref{eqn:aic}, 
    is that of \gls{bic}, 
\begin{equation}
    \label{eqn:bic}
    BIC_i = k \ \ln(n) - 2 \tll_{i},
\end{equation}
    where $k, \tll_{i}$ are as defined above and $n$ is the number of samples.
Analagously to Akaike weights, 
    \emph{Bayes weights} as proposed in \S7.7 of \cite{friedman2001elements}, are given by

\begin{equation}
    w_i^B = \exp\left( - \frac{BIC_i}{2}  \right).
\end{equation}

\gls{bic} introduces a stronger penalty in the number of parameters in each model, when compared with \gls{aic}, i.e. a strong statistical evidence is needed to justify any additional parameters. 
Again, this may be overly cumbersome for our use case:
    with such a relatively small number of parameters, 
    the punishment is disproportionate; 
    moreover since we are trying to uncover physical interactions, 
    we do not necessarily want to suppress models merely for 
    their cardinality, since this might result in favouring 
    simple models which do not capture the physics.  
As with Akaike weights, then, we opt instead for a simpler objective function,
\begin{equation}
    \label{eqn:bic_fitness}
    g_i^B = \left( \frac{1}{BIC_i}\right)^2.
\end{equation}

\subsection{Bayes factor points}
\label{sec:bf_points}

A cornerstone of model selection within \gls{qmla} is the calculation of \gls{bf} (see \ref{sm:bayes_factors}). 
We can compute the pairwise \gls{bf} between two candidate models, $\bij$, according to \cref{eqn:bf_succinct}.
$\bij$ can be based on $\expset_v$, but can also be calculated from $\expset_i \cup \expset_j$: 
    this is a strong advantage since the resulting insight (\cref{eqn:bf_cases}) is based on 
    experiments which were bespoke to both $\hi, \hj$. 
As such we can be confident that this insight accurately points us to the stronger of two candidate models. 
\par 

We can utilise this facility by simply computing the \gls{bf} between all 
    pairs of models in a set of $N_m$ candidates $\{\hi\}$, 
    i.e. compute ${N_m \choose 2}$ \gls{bf}s. 
Note that this is computationally expensive: 
    in order to train $\hi$ on $\expset_j$ requires a further $\left| \expset_j \right|$ experiments, 
    each requiring $N_P$ particles\footnote{Caveat the reduction in overhead outlined in \ref{sm:bayes_factors}.}, 
    where each particle corresponds to a unitary evolution and therefore the caluclation of a matrix exponential. 
The combinatorial scaling of the model space is then quite a heavy disadvantage. 
However, in the case where all pairwise \gls{bf} are performed, 
    we can assign a point to $\hi$ for every comparison which favours it. 
\begin{equation}
    \label{eqn:bf_points}
    g_i^p = \sum_j b_{ij}, \ \ \ \ b_{ij} = 
        \begin{cases}
            1, \ \ \ \ \bij > 1 \\
            0, \ \ \ \ \text{otherwise}
        \end{cases}
\end{equation}
This is a straightforward mechanism, but is overly blunt
    because it does not account for the strength of the evidence
    in favour of each model. 
For example, a dominant model will receive only a slightly higher selection probability 
    than the second strongest, even if the difference between them was $\bij = 10^{100}$. 
Further, the unfavourable scaling make this an expensive method. 

\subsection{Ranking}\label{sec:bf_ranking}
Related to \ref{sec:bf_points}, we can rank models in a generation based on their number of \gls{bf} points.
\gls{bf} points are assigned as in \cref{eqn:bf_points}, 
    but instead of corresponding directly to fitness, 
    we assign models a rank $R$, 
    i.e. the model with highest $g_i^p$ gets $R=1$, 
    and the model with $n^{th}$--highest $g_i^p$ gets $R=n$. 
Note here we truncate, meaning we remove the worse-performing models and retain only $N_m^{\prime}$ models, 
    before calculating $R$.
This is because computing $R$ using all $N_m$ models results in less distinct selection probabilities. 

\begin{equation}
    \label{eqn:ranking}
    g_i^R = \frac{N_m^{\prime}-R+1}{\sum\limits_{n=1}^{N^{\prime}_m} n},
\end{equation}
    where $R$ is the rank of $\hi$ and $N_m^{\prime}$ is the number of models retained after truncation.

\subsection{Residuals}\label{sec:residuals}
At each experiment, $N_P$ particles are compared against a single experimental datum, $d$. 
For consistency with QInfer \cite{qinfer-1_0} -- on which \gls{qmla}'s code base extends --
    we call the expectation value for the system $\Pr_Q(0)$, 
    and that of each particle $\Pr_p(0)$. 
When evolving against a time-independent Hamiltonian, $\Pr_Q(0) = \left| \bra{\psi} e^{-i\ho t} \ket{\psi} \right|^2$, 
    but this can be changed to match given experimental schemes, 
    e.g. the Hahn-echo sequence applied in \cite{gentile2020learning}. 
By definition, the datum $d$ is the binary outcome of the measurement on the system under experimental conditions $e$.
That is, $d$ encodes the answer to the question: 
    after time $t$ under Hamiltonian evolution, did $Q$ project onto 
    the basis state labelled $\ket{0}$ (usually the same as the input probe state $\ket{\psi}$)?
However, in practice we often have access also to the expectation value, 
    i.e. rather than a binary value, a number representing the probability that 
    $Q$ will project on to $0$ for a given experiment $e$, $\Pr_Q(0 | e)$. 
Likewise, we can simulate this quantity for each particle, $\Pr_p(0 | e)$. 
This allows us to calculate the \emph{residual} between the system and individual particles, $r_p^e$, 
    as well as the mean residual across all particles in a single experiment $r^e$:
\begin{align}
    \label{eqn:particle_residual}
    \begin{split}
    r_p^e & = \left| Pr_Q(0 | e) - Pr_p(0 | e) \right|  \\
    r^e &= \underset{p}{\text{mean}} \{r_p^e\}
    \end{split}
\end{align}

Then, for a single experiment $e$, we wish to characterise the quality of the entire parameter distribution $P({\alpha})$. 
We can do this by taking the mean residual across all particles, $\bar{r}^e_p$. 
\par 
Residuals capture how closely the particle distribution reproduced the dynamics from $Q$:
    $r^{e}_{p} = 0$ indicates perfect preditiction, while $r^e_p=1$ is completely incorrect. 
We can therefore maximise the quantity $1-r$ to find    the best model, 
    using the \gls{of}
\begin{equation}
    \label{eqn:residual_fitness}
    g^r_i = \left| 1 - \underset{e \in \expset}{\text{mean}}\{ {r^e} \}\right|^2.
\end{equation}
\par     
This \gls{of} can be thought of in frequentist terms 
    as similar to the residual sum of squares,
    although instead of summing the residual squares, we average to ensure $0 \leq r \leq 1$. 
$g_i^r$ encapsulates how well a candidate model can reproduce dynamics 
    from the target system, as a proxy for whether that candidate describes the system. 
This is not always a safe figure of merit: 
    in most cases, we do not expect parameter learning to perfectly optimise $\al_i$. 
Reproduced dynamics alone can not capture the likelihood that $\hi=\ho$. 
However, this \gls{of} provides a useful test for \gls{qmla}'s \gls{ga}:
    by simulating the case where parameters \emph{are} learned perfectly, 
    such that we know that $g_i^r$ truly represents the ability of $\hi$ to 
    simulate $\ho$, then this \gls{of} guarantees to promote  the strongest models,
    especially given that $\hi=\ho \implies r_p^e=0 \ \forall \ \{e, p\}$. 
In realistic cases, however, the non-zero residuals -- even for 
    strong $\hi$ -- may arise from imperfectly learned parameters,
    rendering the usefulness of this \gls{of} uncertain. 
Finally, it does not account for the cardinality, $k$, of the candidate models, 
    which could result in favouring severely overfitting models in order to gain marginal improvement 
    in residuals, which all machine learning protocols aim to avoid in general.

\subsection{Bayes-factor-enhanced Elo-ratings}\label{sec:elo}
A popular tool for rating individual competitors in sports and games is the \emph{Elo rating} scheme, 
    e.g. used to rate chess players and soccer teams \cite{elo1978rating, fifa_elo}, 
    also finding application in the study of animal hierarchies \cite{neumann2011assessing}. 
Elo ratings allow for evaluating relative quality of individuals 
    based on incomplete pairwise competitions, 
    e.g. despite two soccer teams having never played against each other before, 
    it is possible to quantify the difference in quality between those teams, 
    and therefore to predict a result in advance \cite{hvattum2010using}. 
We recognise a parallel with these types of competition by noting that
    in our case, we similarly have a pool of individuals (models), 
    which we can place in direct competition, and quantify the outcome through \gls{bf}. 

\par 
Elo ratings are transitive: given some interconnectivity in a generation, 
    we need not compare \emph{every} pair of models in order to 
    make meaningful claims about which are strongest;
    it is sufficient to perform a subset of comparisons, 
    ensuring each individual is tested robustly. 
We can take advantage of this transitivity to reduce the combinatorial overhead 
    usually associated with computing bespoke \gls{bf} between all models 
    (i.e. using their own training data $\expset_i$ instead of a generic 
    $\expset_v$).
In practice, we map models within a generation to nodes on a graph,
    which is then sparsely connected. 
In composing the list of edges for this graph, 
    we primarily prioritise each node having a similar number of edges,
    and secondarily the distance between any two nodes. 
For example, with 14 nodes we overlay edges such that each node 
    is connected with 5,6 or 7 other nodes, 
    and all nodes at least share a competitor in common. 
\par 

The Elo rating scheme is as follows: 
    upon creation, $\hi$ is assigned a rating $R_i$; 
    every comparison with a competitor $\hj$ results in $\bij$; 
    $R_i$ is updated according to the known strength of its competitor, $R_j$, 
    as well as the result $\bij$. 
The Elo update ensures that winning models are rewarded 
    for defeating another model, 
    but that the extent of that reward reflects the quality of its opponent. 
As such, this is a fairer mechanism than \gls{bf} points, 
    which award a point for every victory irrespective of the opposition:
    if $\hj$ is already known to be a strong or poor model, 
    then $\Delta R_i$ proportionally changes the credence we assign to $\hi$. 
It achieves this by first computing the \emph{expected} result of a given comparison
    with respect to each model, based on the current ratings, 

\begin{subequations}
    \label{eqn:elo_expected_score}
    \begin{equation}
        E_i 
        = \frac{1}{1 + 10^{\frac{R_j - R_i}{400}}} ;
    \end{equation}
    \begin{equation}
        E_i + E_j = 1,
    \end{equation}    
\end{subequations}

Then, we find the binary \emph{score} from the perspective of each model:
\begin{equation}
    \label{eqn:elo_score}
    \begin{cases}
        B_{ij} > 1 & \Rightarrow \ S_i = 1; \ S_j =0  \\
        B_{ij} < 1 & \Rightarrow \ S_i = 0; \ S_j = 1 \\
    \end{cases}
\end{equation}
which is used to determine the change to each model's rating:
\begin{equation}
    \label{eqn:elo_delta_r}
    \Delta R_i = \eta \times \left( S_i - E_i \right).
\end{equation}

An important detail is the choice of $\eta$, i.e. the \emph{weight} of the 
    change to the models' ratings. 
In standard Elo schemes this is a fixed constant, 
    but here -- taking inspiration from soccer ratings where $\eta$ is the number of goals 
    by which one team won -- we weight the change by the strength of our belief in the outcome: 
    $\eta \propto \left| \bij \right|$.
That is, similarly to the interpretation of \cref{eqn:bf_cases}, 
    we use the evidence in favour of the winning model to transfer points from the loser to the winner,
    albeit we temper this by instead using $\eta = log_{10}(B_{ij})$, since \gls{bf} can give very large numbers. 
In total, then, following the comparison between models $\hi, \hj$, we can perform the Elo rating update
\begin{equation}
    \label{eqn:elo_update}
    R_i^{\prime} = R_i + \text{log}_{10}(B_{ij}) \left(S_i - \frac{1}{1 + 10^{\frac{R_j - R_i}{400}}}\right).
\end{equation}
Again, this procedure is easiest to understand by following the example in \cref{table:elo_eg}. 

\begin{table}
    \centering
    \begin{tabular}{|cc|c|c|c|c|c|c|c|}
\hline
                        &             &                                    $R_i$ &                                    $E_i$ &                                    $S_i$ &                                 $B_{ij}$ &                       $\log_{10}(B_{ij})$ &                             $\Delta R_i$ &                           $R_i^{\prime}$ \\
 & Model &                                          &                                          &                                          &                                          &                                          &                                          &                                          \\
\hline
\hline \multirow{2}{*}{$\hat{H}_a > \hat{H}_b$} & $\hat{H}_a$ &                                     1000 &                                     0.76 &                                        1 &                                   1e+100 &                                      100 &                                     0.24 &                                   1024.0 \\
                        & $\hat{H}_b$ &                                      800 &                                     0.24 &                                        0 &                                   1e-100 &                                      100 &                                    -0.24 &                                    776.0 \\
\cline{1-9}
\hline \multirow{2}{*}{$\hat{H}_b > \hat{H}_a$} & $\hat{H}_a$ &                                     1000 &                                     0.76 &                                        0 &                                   1e-100 &                                      100 &                                    -0.76 &                                    924.0 \\
                        & $\hat{H}_b$ &                                      800 &                                     0.24 &                                        1 &                                   1e+100 &                                      100 &                                     0.76 &                                    876.0 \\
\hline
\end{tabular}

    \caption{
        Example of Elo rating updates. We have two models, where $\ha$ is initially stronger than $\hb$, 
            and we show the result of a strong result in favour of each model. 
        Note to achieve $B_{ij} = 10^{100} = e^{\tll_{i} - \tll_{j}} \implies \tll_{i} - \tll_{j} = \ln(10^{100}) \approx 7$. 
        In the case where $\ha$ defeats $\hb$, because this was so strongly expected given their initial ratings, 
            the reward for $\ha$ (and cost to $\hb$) is relatively small, 
            compared with the case where -- contrary to prediction -- $\hb$ defeats $\ha$. 
    }
    \label{table:elo_eg}
\end{table}

\par 
Finally, it remains to select the starting rating $R_i^0$ to assign models upon creation. 
Although this choice is arbitrary, it can have a strong effect on the progression of the algorithm. 
Here we impose details specific to the \gls{qmla} \gls{ga}: 
    at each generation we admit the top two  models automatically 
    for consideration in the next generation, 
    such that strongest models can stay alive in the population and ultimately win. 
These are called \emph{elite} models, $\hat{H}_e^1, \hat{H}_e^2$. 
This poses the strong possibility for a form of generational wealth:
    if elite models have already existed for several generations, 
    their Elo ratings will be higher than all alternatives by defintion. 
Therefore by maintaining a constant $R_i^0$, 
    i.e. a model born at generation 12 gets the same $R_i^0$ as $\hat{H}_e^1$ -- which was 
    born several generations prior and has been winning \gls{bf} comparisons ever since --  
    we bias the \gls{ga} to continue to favour the elite models. 
Instead, we would prefer that newly born models can overtake the Elo rating of elite models. 
At each generation, all models -- including the elite models from the prvious generation -- 
are assigned the same initial rating. 
This ensures that we do not unfairly favour elite models: 
    new models have a reasonably chance of overtaking the elite models.
    

\par 

Given the arbitrary scaling of the Elo rating scheme, 
    and in order to derive a meaningful selection probability, 
    we ought to ground the raw Elo rating somehow at each generation $\mu$. 
We do this by subtracting the lowest rating among the entertained models, $R_{\textrm{min}}^{\mu}$.
This serves to ensure the range of $R_i$ remaining is defined only by 
    the difference between models as assessed within $\mu$: 
    a very strong model might have much higher $R_i$ than its contemporaries, 
    but that difference was earned exclusively by comparison within $\mu$, 
    so it deserves higher fitness. 
We perform this step before truncation, so that the reamining models post-truncation
    all have non-zero fitness. 
Finally, then, we name this \gls{of} the \emph{Bayes-factor enhanced Elo rating}, 
    whereby the fitness of each model is attained directly from its rating
    after undergoing Elo updates in the current generation minus the minimum rating of any model 
    in the same generation $\mu$,

\begin{equation}
    \label{eqn:elo_fitness}
    g_i^E = R_i^{\mu} - R_{\textrm{min}}^{\mu}.
\end{equation}
The advantage of this \gls{of} is that it gives a meaningful value on the absolute quality of every model, 
    allowing us to determine the strongest, and importantly to find the relative strength between models. 
Further, it exploits bespoke \gls{bf}, i.e. based on the considered models' 
    individually designed $\expset_i$,
    removing the impetus to design $\expset_v$ which can
    evaluate models definitively. 
One disadvantage is that it does not explicitly punish models based 
    on their cardinality, 
    however this feature is partially embedded by adopting \gls{bf} for the comparisons, 
    which are known to protect against overfitting.

\subsection{Choice of objective function}
\begin{figure}
    \centering
    \includegraphics{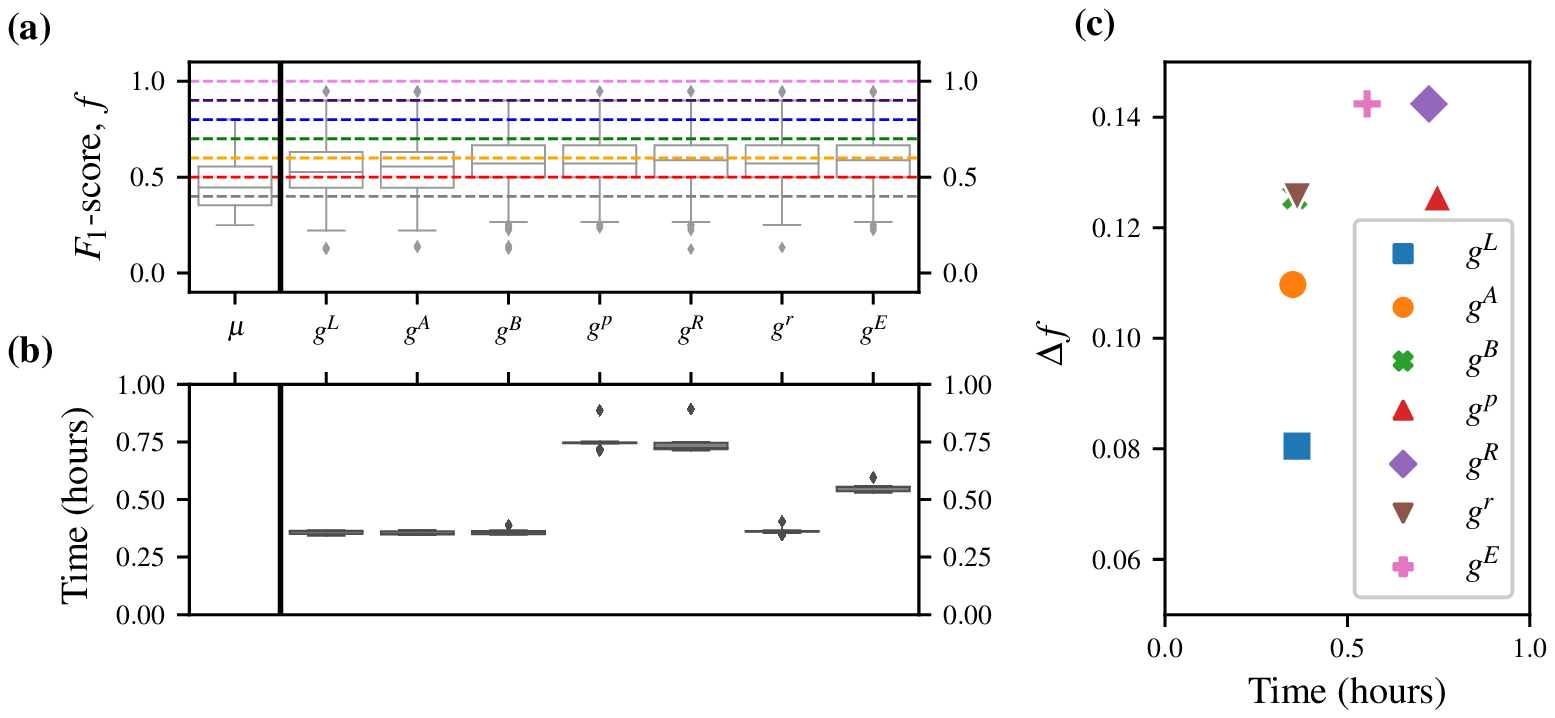}
    \caption{
        Comparison between proposed \gls{of}. 
        Each \gls{of} trains the same initial generation of $N_m=28$ models with resources
        $N_E=500, N_P=2000$, and then design a new set of $N_m$ models through 
        the same roulette strategy, such that the only difference between \gls{of}'s output 
        is how they assign selection probability.
        We run each \gls{of} 25 times for the same target system as used in the main text, 
            i.e. a $4$--qubit Heisenberg--XYZ model. 
        \textbf{(a)} shows the box--plot of new models' \fs \ $f$, 
            where the median and inter--quartile ranges are indicated by the boxes,
            as well as those of the initial generation $\mu$ centered on $f_{\mu}=0.45$.
            We mark $f=\{0.4, 0.5, ..., 1.0\}$ for ease of interpretation. 
        \textbf{(b)} shows box--plots of the time taken to compute the single generation in each case.
        In \textbf{(c)} we report the difference between the median $f$ among the 
            newly proposed models from $f_{\mu}$, $\Delta f$,
            plotted against the time to achieve the result. 
    }
    \label{fig:obj_fnc_comparison}
\end{figure}

Having proposed a series of possible objective functions, 
    we are now in a position to analyse which of those are most appropriate for \gls{qmla}. 
Recall from \ref{sec:f_score} the figure of merit we use for models, \fs, 
    which we will use to distinguish between the outputs of each \gls{of}.
We test each \gls{of} by the quality of models they produce: 
    we train the same batch of models in each case and allow each \gls{of} 
    to compute the selection probabilities and therefore direct the design 
    of the next generation of models. 
We input the same set of $N_m=28$ models, 
    randomly selected from the same model space used in the main text, also using the same target model,
    and we plot the distribution of \fs \ that each \gls{of} produces in \cref{fig:obj_fnc_comparison}. 
We also account for the time taken in each case: 
    we report the time to train and evaluate the single generation on a 16--core node.
Overall, then, we observe how Bayes-factor-enhanced Elo ratings balance well a satisfactory outcome with economy of computational resources. Therefore, we adopted it for the analyses in the main text.
We strongly emphasise, however, that the performance of each objective function
    can vary under alternative conditions, and therefore similar analysis may 
    be warranted for future applications. 
For instance, if $t_{max}$ is known to be small, 
    in smaller model spaces, using $g^r$ results in higher success rates.

\glsreset{of}

\subsection{Use of $F_1$-score as proxy for model quality}\label{sm:f_score_justification}
We have relied on the $F_1$-score to capture the success of \gls{qmla} throughout the main text, as defined in \ref{sec:f_score}. 
We did not use $F_1$-score in the generation of models -- it is used only to assess the performance of \gls{qmla}. 
While it is not guaranteed that models of higher $f$ should be superior in predicting the dynamics of a quantum system, here we demonstrate that, usually, in a competition of trained models, the favoured model indeed holds the higher $F_1$-score. 
This serves as motivation for the model generation strategies employed: we expect that comparisons between models will -- in the majority, but not all cases -- yield that model which overlaps more with the target model. 
Therefore by iterating the procedure, on average \gls{qmla} will uncover more terms which are \emph{present} in the set of true terms, $\mathcal{T}_0$. 
\par

\begin{figure}
    \centering
    \includegraphics{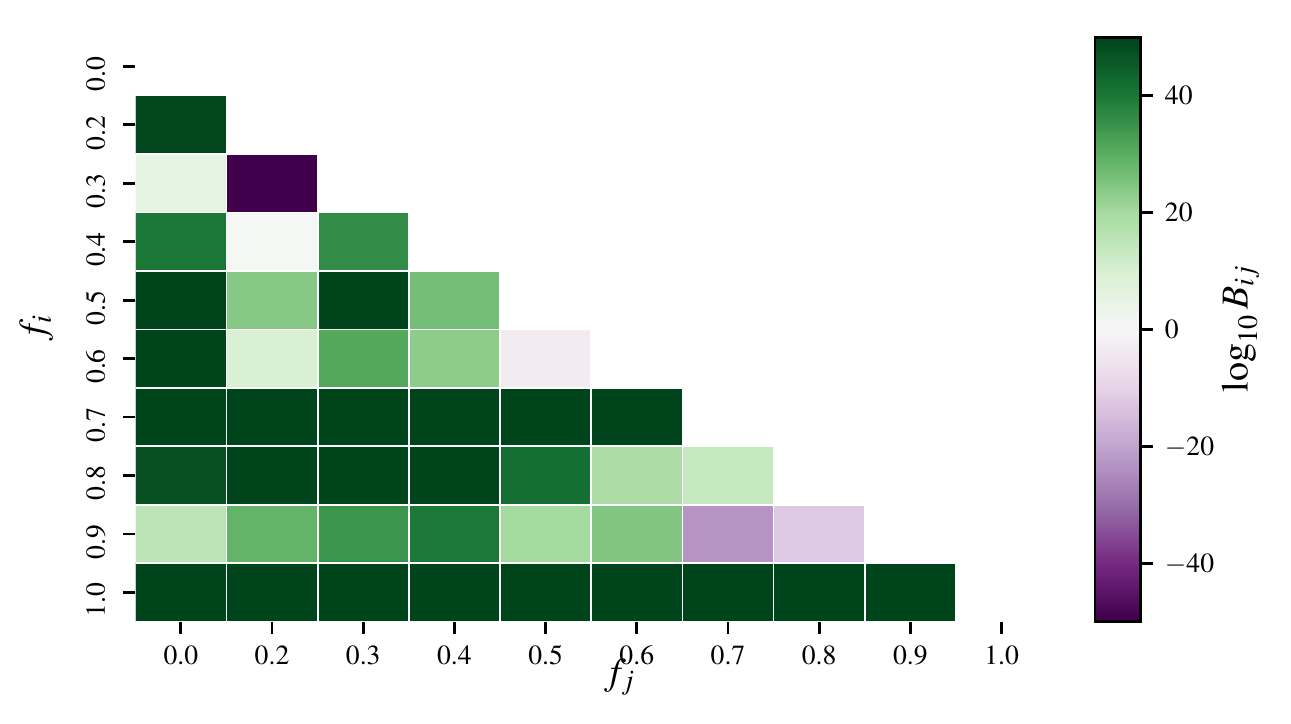}
    \caption{
        Pairwise \acrlong{bf} by $F_1$-score. 
        Each tile shows $\log_{10}\bij$ against the $F_1$-score of the two candidates $\hi$ ($f_i$ on the $y$-axis) 
        and $\hj$ ($f_j$ on the $x$-axis).
        $\log_{10}\bij > 0 $ shown in green ($\log_{10}\bij < 0 $ shown in purple) 
            indicates statistical evidence that $\hi$ ($\hj$) 
            is the better model with respect to the observed data.
        Visualisation is curtailed to $\log_{10} \bij = \pm 50$.
    }
    \label{fig:bayes_factors_by_f_score}
\end{figure}

\cref{fig:bayes_factors_by_f_score} shows $B_{ij}$, the \gls{bf} comparisons between models of varying $F_1$-score, $f_i$ and $f_j$.
In almost all cases, $B_{ij} > 10^{50}$ when $f_i > f_j$. 
Recall that \gls{bf} are a statistical measure of the relative ability of two models to explain the same observed dataset, \ref{sm:bayes_factors}.
\cref{eqn:bf_succinct} shows the \gls{bf} as the exponential of the difference between the log-likelihoods for $\hi$ and $\hj$: the log-likelihood is a cumulative measure of the goodness-of-fit of each model, therefore the \gls{bf} indicates which model best captures the dynamics of $Q$, since the dataset against which the \gls{bf} is computed are measurements of the system after evolving for some time. 
Then, we can see from \cref{fig:bayes_factors_by_f_score} that in a pairwise comparison, the model with higher $F_1$-score is usually favoured by the \gls{bf}, and therefore fits the data more strongly, confirming our expectation that models which share more terms with $\ho$ are better capable of reproducing the dynamics of $Q$. 

    \section{Algorithm details}\label{sm:algorithm}
Here we list pseudocode for some of the core algorithms and subroutines discussed regularly throughout this work. 
Alg \ref{alg:qmla} shows the overall \gls{qmla} algorithm, 
    which is simplified greatly to a loop over model searches for each \gls{es}. 
The model search itself is listed in \cref{alg:model_search},
    which contains calls to subroutines for model learning (\gls{qhl}, \cref{alg:qhl}), 
    branch evaluation (which can be based upon \gls{bf}, \cref{alg:bayes_factor})
    and centers on the generation of new models, \cref{alg:generate_models}.

\begin{algorithm}
    \caption{Quantum Model Learning Agent}
    \label{alg:qmla}
    \DontPrintSemicolon
    \KwIn{ $Q$ \tcp*[1]{some physically measurable or simulateable quantum system}}
    \KwIn{ $\mathbb{S}$ \tcp*[1]{set of exploration strategies}}\;

    \KwOut{$\hat{H}^{\prime}$ \tcp*[1]{champion model}}\;
    
    $\mathbb{H}_c \gets \{\}$\;   
    \For{$S \in \mathbb{S}$ }{
        $\hat{H}_{S}^{\prime} \gets$ \ttt{model\_search(Q, S)} 

        $\mathbb{H}_c \gets \mathbb{H}_c \cup  \{\hat{H}_{S}^{\prime}\}$ \tcp*[1]{add ES champion to collection}
    }
    $\hat{H}^{\prime} \gets $ \ttt{final\_champion($\mathbb{H}_c$)}\;
    return $\hat{H}^{\prime}$

\end{algorithm}

\begin{algorithm}
    \caption{ES subroutine: \ttt{model\_search}}
    \label{alg:model_search}
    \DontPrintSemicolon
    \KwIn{ $Q$ \tcp*[1]{some physically measurable or simulateable quantum system}}
    \KwIn{ $S$ \tcp*[1]{Exploration strategy: collection of rules/subroutines}}\;

    \KwOut{$\hat{H}_{S}^{\prime}$ \tcp*[1]{Exploration strategy's nominated champion model}}\;
     
    $\nu \gets \{\}$

    $\mathbb{H}_c \gets \{\}$\;   
    \While{\rm{no termination occurs for} \ttt{S}}
    {
        $\mu \gets $ \ttt{generate\_models($\nu$)} \rm{according to} \ttt{S} \tcp*[1]{e.g. \cref{alg:generate_models}}\; 

        \For{ $\hi \in \mu$}{
            $\hi^{\prime} \gets $ \ttt{train($\hi$)} \tcp*[1]{e.g. \cref{alg:qhl}}
        }
        $\nu \gets$ \ttt{evaluate($\mu$)} \tcp*[1]{e.g. pairwise via \cref{alg:bayes_factor}}

        $\hat{H}_{c}^{\mu} \gets $ \ttt{branch\_champion($\nu$)} \tcp*[1]{use $\nu$ to select a branch champion}

        $\mathbb{H}_c \gets \mathbb{H}_c \cup \{\hat{H}_c^{\mu}\}$ \tcp*[1]{add branch champion to collection}
    }

    $\hat{H}_{S}^{\prime} \gets $ \ttt{nominate\_champion($\mathbb{H}_c$)}\;
    return $\hat{H}_{S}^{\prime}$

\end{algorithm}

\begin{algorithm}
    \caption{ES subroutine: \ttt{generate\_models} (example: genetic algorithm)}
    \label{alg:generate_models}
    \DontPrintSemicolon
    \KwIn{ $\nu$ \tcp*[1]{information about models considered to date}}\;
    \KwIn{ $g(\hi)$ \tcp*[1]{objective function}}\;

    \KwOut{$\mathbb{H}$ \tcp*[1]{set of models}}\;
    
    $ N_m = \left| \nu \right| $ \tcp*[1]{number of models}

    \For{$\hi \in \nu$}{
       $g_i \gets g(\hi)$ \tcp*[1]{model fitness via objective function}
    }
    $r \gets $ \ttt{rank($\{ g_i \}$)} \tcp*[1]{rank models by their fitness}
    $\mathbb{H}_t \gets $ \ttt{truncate($r, \frac{N_m}{2}$)} \tcp*[1]{truncate models by rank: only keep $\frac{N_m}{2}$}
    $ s \gets $ \ttt{normalise($\{g_i\}$)} $\forall \hi \in \mathbb{H}_t$ \tcp*[1]{normalise remaining models' fitness}

    $\mathbb{H} = \{\}$ \tcp*[1]{new batch of chromosomes/models}

    \While{ $\left| \mathbb{H} \right| < N_m$ }{
        $p_1, p_2 = $ \ttt{roulette($s$)} \tcp*[1]{use $s$ to select two parents via roulette selection}
    
        $c_1, c_2$ = \ttt{crossover($p_1, p_2$)} \tcp*[1]{produce offspring models}

        $c_1, c_2$ = \ttt{mutate($c_1, c_2$)} \tcp*[1]{probabilistically mutate}

        $\mathbb{H} \gets \mathbb{H} \cup \{ c_1, c_2 \} $ \tcp*[1]{add new models to batch}

    }


    return $\mathbb{H}$ 

\end{algorithm}

\begin{algorithm}
    \caption{Quantum Hamiltonian Learning}
    \label{alg:qhl}
    \DontPrintSemicolon
    \KwIn{ $Q$ \tcp*[1]{some physically measurable or simulatable quantum system, described by $\ho$}}

    \KwIn{$\hat{H}_i$ \tcp*[1]{Hamiltonian model attempting to reproduce data from $\hat{H}_0$ }}
    \KwIn{$P_{\al}$ \tcp*[1]{probability distribution for $\al = \al_0 $}}
    \KwIn{$N_E$ \tcp*[1]{number of epochs to iterate learning procedure for}}
    \KwIn{$N_P$ \tcp*[1]{number of particles to draw from $P_{\al}$}}
    \KwIn{\ttt{$\Lambda(P_{\al})$} \tcp*[1]{Heuristic algorithm which designs experiments}}
    \KwIn{\ttt{RS($P_{\al}$)} \tcp*[1]{Resampling algorithm for redrawing particles}}

    \KwOut{$\al^{\prime}$ \tcp*[1]{estimate of Hamiltonian parameters}}\;

    Sample $N_P$ times from $P_{\al} \gets \mathcal{P}$ \tcp*[1]{particles}\;
    
    \For{$ e \in  \{1 \rightarrow N_E\}$ }{ 
        
        $t, \ket{\psi} \gets \Lambda(P_{\al})$ \tcp*[1]{design an experiment}
        
        \For{
            $p \in \mathcal{P}$ 
        }{
            Retrieve particle $p \gets \al_p$
            
            Measure Q at $t \gets d$ \tcp*[1]{datum}
            
            $| \bra{\psi} e^{-iH(\al_p)t} \ket{\psi} |^2 \gets E_p$ \tcp*[1]{expectation value}
            
            $Pr(d|\al_p; t) \gets l_p$ \tcp*[1]{likelihood, computed using $E_p$ and $d$}
            
            $w_p \gets w_p \times l_p $ \tcp*[1]{weight update}
        }
        \If{
            $ 1 / \sum_p w_p^2 < N_P/2$
            \tcp*[1]{check whether to resample (are weights too small?)}
        }{
            $P_{\al} \gets \ttt{RS}(P_{\al})$
            \tcp*[1]{Resample according to provided resampling algorithm}
            Sample $N_P$ times from $P_{\al} \gets \mathcal{P}$
        }
    }
    
    \ttt{mean}($P_{\al}) \gets \vec{\alpha}^{\prime}$\;
    
    return $\al^{\prime}$
     
\end{algorithm}

\begin{algorithm}
    \caption{Bayes Factor calculation}
    \label{alg:bayes_factor}
    \DontPrintSemicolon
    \KwIn{ $Q$ \tcp*[1]{some physically measurable or simulateable quantum system.}}
    \KwIn{ $\hat{H}_j^{\prime}, \hat{H}_k^{\prime}$ \tcp*[1]{Hamiltonian models after QHL (i.e. $\al_j, \al_k$ already optimised), on which to compare performance.}}
    \KwIn{ $\expset_j, \expset_k$ \tcp*[1]{experiments on which $\hat{H}_j^{\prime}$ and $\hat{H}_k^{\prime}$ were trained during QHL.}}
    
    \KwOut{$B_{jk}$ \tcp*[1]{Bayes factor between two candidate Hamiltonians}}

    $\expset = \{ \expset_j \cup \expset_k $\} \;
    \For{$\hat{H}_i^{\prime} \in  \{\hat{H}_j^{\prime}, \hat{H}_k^{\prime}$ \} }{
        $\tll_{i} = 0$ \tcp*[1]{total log-likelihood of $\hat{H}_i$} 
        \For{$(t, \ket{\psi}) \in \expset$}{

            Measure Q at $t \gets d$ \tcp*[1]{datum for time $t$}
            
            Compute $| \bra{\psi} e^{-i \hat{H}_i^{\prime} t} \ket{\psi} |^2 \gets E_p$  \tcp*[1]{expectation value for $\hat{H}_i^{\prime}$ at $t$} 
            
            Compute $Pr(d | \hat{H}_i, t)$ \tcp*[1]{from Bayes' rule, using $d, E_p$}
            
            $log \left( Pr(d | \hat{H}_i, t) \right) \gets l$ \tcp*[1]{likelihood of observing datum $d$ for $\hat{H}_i^{\prime}$ at $t$}
            
            $\tll_{i}+ l \gets  \tll_{i}$ \tcp*[1]{add this likelihood to total log-likelihood }
        }
    }
    
    $exp\left( \tll_{j} - L_k \right) \gets B_{jk}$  \tcp*[1]{Bayes factor between models}\;
    return $B_{jk}$
\end{algorithm}

    \clearpage


    \clearpage
    \bibliography{references_main}

\end{document}


    \date{\today}
    \newacronym{qmla}{QMLA}{quantum model learning agent}
\newacronym{qhl}{QHL}{quantum Hamiltonian learning}
\newacronym{bf}{BF}{Bayes factor}
\newacronym{nisq}{NISQ}{noisy intermediate scale quantum}
\newacronym{ml}{ML}{machine learning}
\newacronym{qml}{QML}{quantum machine learning}
\newacronym{nv}{NV}{nitrogen-vacancy}
\newacronym{ga}{GA}{genetic algorithm}
\newacronym{gr}{GR}{Growth Rule}
\newacronym[plural=ESs, firstplural=exploration strategies (ESs)]{es}{ES}{exploration strategy}
\newacronym{et}{ET}{exploration tree}
\newacronym{fh}{FH}{Fermi-Hubbard}
\newacronym{fhm}{FHM}{Fermi-Hubbard model}
\newacronym{im}{IM}{Ising model}
\newacronym{hm}{HM}{Heisenberg model}
\newacronym{bfeer}{BFEER}{Bayes factor enhanced Elo ratings}
\newacronym{ges}{GES}{genetic exploration strategy}
\newacronym{smc}{SMC}{sequential Monte Carlo}
\newacronym{tll}{TLL}{total log-likelihood}
\newacronym{aic}{AIC}{Akaike information criterion}
\newacronym{aicc}{AICC}{Akaike information criterion corrected}
\newacronym{bic}{BIC}{Bayesian information criterion}
\newacronym{ll}{LL}{log-likelihood}
\newacronym{of}{OF}{objective function}
\newacronym{tp}{TP}{true positives}
\newacronym{tn}{TN}{true negatives}
\newacronym{fp}{FP}{false positives}
\newacronym{fn}{FN}{false negatives}
\newacronym{iqle}{IQLE}{interactive quantum likelihood estimation}

    \clearpage
    \onecolumn

    \section*{Supplementary Material}
    \renewcommand \thesection{\arabic{section}}
    \renewcommand*\thefigure{SM\arabic{figure}}  
    \renewcommand*\thetable{SM\arabic{table}}  
    \renewcommand{\theequation}{SM\arabic{equation}}
    \setcounter{figure}{0}    
    \setcounter{equation}{0}
    \appendixpageoff
    \appendixtitleoff
    
    \glsresetall
    \begin{appendices}
        \section{Glossary}\label{sm:glossary}

        \section{Subroutines}\label{sm:subroutines}
        \subfile{../content/subroutines}

        \section{Exploration Strategies}\label{sm:exploration_strategies}
        \subfile{../content/exploration_strategies.tex}

        \section{Objective Functions}\label{sm:objective_fncs}
        \subfile{../content/objective_fncs}
        
        \section{Algorithm details}\label{sm:algorithm}
        \subfile{../content/pseudocode.tex}
        
        \clearpage
    \end{appendices}
    \renewcommand{\appendixname}{Supplementary Material}

    \printbibliography

    \clearpage